\documentclass[aps,onecolumn,showpacs,superscriptaddress,floatfix]{revtex4}
\usepackage{graphicx}
\usepackage{psfrag}
\usepackage{color}

\usepackage{amsmath}
\usepackage{amssymb}   
\usepackage{amstext}
\usepackage{amsthm}
\usepackage{amsfonts}
\usepackage{amsxtra}

\begin{document}

\newcommand{\ket}[1]{\left\vert #1\,\right\rangle}
\newcommand{\ketrc}{\left\vert\textsc{rc}\,\right\rangle}
\newcommand{\bra}[1]{\left\langle #1\,\right\vert}
\newcommand{\bbra}[1]{\left\langle\left\langle #1\,\right\vert\right.}
\newcommand{\brarc}{\left\langle\,\textsc{rc}\right\vert}
\newcommand{\braket}[2]{\left\langle #1\,\vert #2\,\right\rangle}
\newcommand{\bbraket}[2]{\left\langle\left\langle #1\,\vert #2\,\right\rangle\right.}
\newcommand{\de}{\partial}
\newcommand{\eps}{\varepsilon}
\newcommand{\tr}{\operatorname{\mathrm{Tr}}}
\newcommand{\re}{\mathrm{Re}}
\newcommand{\im}{\mathrm{Im}}
\newcommand{\R}{\mathbb{R}}
\newcommand{\eav}[1]{\left\langle #1\right\rangle}
\newcommand{\beq}{\begin{equation}}
\newcommand{\eeq}{\end{equation}}
\newcommand{\ben}{\begin{eqnarray}}
\newcommand{\een}{\end{eqnarray}}
\newcommand{\bea}{\begin{array}}
\newcommand{\eea}{\end{array}}
\newcommand{\om}{(\omega )}
\newcommand{\bef}{\begin{figure}}
\newcommand{\eef}{\end{figure}}
\newcommand{\leg}[1]{\caption{\protect\rm{\protect\footnotesize{#1}}}}
\newcommand{\ew}[1]{\langle{#1}\rangle}
\newcommand{\be}[1]{\mid\!{#1}\!\mid}
\newcommand{\no}{\nonumber}
\newcommand{\etal}{{\em et~al }}
\newcommand{\geff}{g_{\mbox{\it{\scriptsize{eff}}}}}
\newcommand{\da}[1]{{#1}^\dagger}
\newcommand{\cf}{{\it cf.\/}\ }
\newcommand{\ie}{{\it i.e.\/}\ }   
\setlength\abovedisplayskip{5pt}
\setlength\belowdisplayskip{5pt}

\title{Opening-Assisted Coherent Transport in the Deep Classical Regime}

\author{Yang \surname{Zhang}}
\affiliation{Department of Physics and Engineering Physics, Tulane University, New Orleans, Louisiana 70118, USA}

\author{G.~Luca \surname{Celardo}}
\affiliation{Dipartimento di Matematica e
Fisica and Interdisciplinary Laboratories for Advanced Materials Physics,
 Universit\`a Cattolica del Sacro Cuore, via Musei 41, I-25121 Brescia, Italy}
\affiliation{Istituto Nazionale di Fisica Nucleare,  Sezione di Pavia, 
via Bassi 6, I-27100  Pavia, Italy}
\affiliation{Instituto de F\'isica, Benem\'erita Universidad Aut\'onoma de Puebla, Apartado Postal J-48, Puebla 72570, Mexico}
\author{Fausto \surname{Borgonovi}}
\affiliation{Dipartimento di Matematica e
Fisica and Interdisciplinary Laboratories for Advanced Materials Physics,
 Universit\`a Cattolica del Sacro Cuore, via Musei 41, I-25121 Brescia, Italy}
\affiliation{Istituto Nazionale di Fisica Nucleare,  Sezione di Pavia, 
via Bassi 6, I-27100  Pavia, Italy}

\author{Lev \surname{Kaplan}}
\affiliation{Department of Physics and Engineering Physics, Tulane University, New Orleans, Louisiana 70118, USA}

\begin{abstract}                
	We study quantum enhancement of transport in open systems in the presence
	of disorder and dephasing. Quantum coherence effects may significantly enhance transport in open systems even in the deep classical regime (where the decoherence rate is greater than the inter-site hopping amplitude), as long as the disorder is sufficiently strong. When the strengths of  disorder and dephasing are fixed, there is an optimal opening strength at which the coherent transport enhancement is optimized. Analytic results are obtained in two simple paradigmatic tight-binding models of large systems: the linear chain and the fully connected network. The physical behavior is also reflected in the FMO photosynthetic complex, which may be viewed as intermediate between these paradigmatic models.
\end{abstract}                                                               
                                                                            
\date{\today}
\maketitle

\section{Introduction}\label{sec:I} 

Since the discovery that quantum coherence may have a functional
role  in biological systems even at room temperature~\cite{photo,photoT,photo2,photo3,schulten}, there
has been great interest in understanding how coherence can be
maintained and used under the influence of different environments with
competing effects. In particular, much recent research has focused on
quantum networks,
due to their relevance to molecular aggregates, such as the J-aggregates~\cite{Jaggr}, natural
photosynthetic systems~\cite{cao1d}, bio-engineered devices for
photon sensing~\cite{superabsorb}, and light-harvesting systems~\cite{sarovarbio}.

Many photosynthetic organisms contain
networks of chlorophyll molecular aggregates in their 
light-harvesting complexes, {\it e.g.} LHI and LHII~\cite{schulten1}. These complexes
absorb light and then transfer the excitations to other
structures or to a central core absorber, the reaction center, where
charge separation, necessary in the
next steps of photosynthesis, occurs.
Exciton transport in biological systems can be interpreted as an energy transfer between chromophores described as  two-level systems. When chromophores are  very close, which for chlorophylls is often less than 10 \AA{}, the interaction between them is manifested in a manner known as exciton coupling. Under low light intensity,  in many 
natural photosynthetic systems or in ultra-precise photon sensors, 
the single-excitation approximation is usually valid. In 
this case the system is
equivalent to a tight binding model where one excitation can hop from
site to site~\cite{cao1d,superabsorb,sarovarbio,fassioli,mukameldeph,mukamelspano}.  

Light-harvesting complexes are subject to the effects of different
environments: $i)$ dissipative,   where the excitation can be
lost; and $ii)$ proteic, which induce static or dynamical disorder.
The efficiency of excitation transfer can be determined only through
a comprehensive analysis of the effects due to the interplay of all those
environments.

Here we consider systems subject to the influence 
of a single decay channel, in the presence of both  static and dynamical
disorder. The decay channel represents coupling to a central core absorber (loss of excitation by trapping).
For many molecular aggregates, the single-channel approximation is appropriate to describe this coupling, modeled for instance by a  semi-infinite one-dimensional lead~\cite{kaplan,giulio,rotterb}. 
The disorder is due to
a protein scaffold, in which
photosynthetic complexes are embedded, that induces 
fluctuations in the site energies. Fluctuations that are slow or fast on the time scale of the dynamics 
are described as static or dynamic disorder, respectively.

Several works in the literature aim to understand 
the parameter regime in which transport efficiency is maximized.  
Some general principles that might be used as a guide to understand
how optimal transport can be achieved have been proposed: 
Enhanced noise assisted transport~\cite{lloyd,deph}, the  Goldilocks principle~\cite{4goldi}, and superradiance
in transport~\cite{srfmo}. 

It is well known that when a quantum system is strongly coupled
to a decay channel, superradiant
behavior may occur~\cite{Zannals}. Superradiance implies the existence of some states with a
cooperatively enhanced decay rate, and is always accompanied by subradiance, the existence of 
states with a cooperatively suppressed decay rate.
Though originally discovered in the context of atomic
clouds interacting with an electromagnetic field~\cite{dicke54}, and in the
presence of many excitations, superradiance 
was soon recognized to be a general phenomenon in open quantum
systems~\cite{Zannals} under the conditions of coherent coupling with
a common decay channel. Crucially,
it can occur in the presence of a
single excitation (the ``super'' of superradiance~\cite{scully}), entailing a purely quantum effect. Most importantly for the
present work, superradiance can have profound effects on transport efficiency in open systems: for example, in a linear chain, the integrated transmission from one end to the other is peaked precisely at the superradiance transition~\cite{kaplan}.

The functional role that superradiance might have in natural
photosynthetic systems 
has been discussed in many publications~\cite{schulten,superabsorb,srlloyd,sr2},
and experimentally observed in molecular aggregates~\cite{Jaggr,vangrondelle}.
Superradiance (or supertransfer) is also thought to play
an important role in the transfer of excitation to the
central core absorber~\cite{schulten},
and its effects on the efficiency of energy transport in
photosynthetic molecular aggregates have recently been analyzed~\cite{srfmo,srrc}.

While superradiance may enhance transport, static disorder is often expected to
hinder it,
since it induces localization~\cite{Anderson}. The relation between superradiance and localization has
been already analyzed in the literature
in different contexts~\cite{mukamelspano,prlT,alberto,cldipole}. Additionally, dynamical disorder (or dephasing) will generally 
destroy cooperativity~\cite{poli}, and hence counteract quantum coherence effects, including superradiance.
On the other hand, dynamical disorder may also enhance efficiency, through
the so-called
noise assisted transport~\cite{lloyd,deph}.

In the deep classical regime where dephasing is stronger than the
coupling between the chromophores, transport in quantum networks can be
described by incoherent master equations with an appropriate choice of 
transition rates. However, the presence of an opening (trapping) introduces a new 
time scale to the system. When the opening strength is large, coherent
effects may be revived even in the deep classical regime.
Here we want to address the following questions: 
$i)$ For which values of the opening strength are coherent effects relevant? $ii)$ Can we enhance transport by increasing the opening, which induces
coherent effects not present in the incoherent model? $iii)$ Under what generic conditions can
coherent effects enhance transport in open quantum systems?

The remainder of the paper is organized as follows. In Sec.~\ref{sec:2}, we present the basic mathematical formalism for analyzing the dynamics of open quantum networks in the presence of both static disorder and dephasing, and define the average transfer time, which measures the transport efficiency in these systems. Then in Sec.~\ref{sec-2site}, we first focus on the two-site model, where all results may be obtained analytically, and determine the regime of dephasing, detuning, and opening strength in which quantum coherent effects enhance quantum transport. Specifically, we show that if the strengths of static and dynamical disorder (detuning and dephasing, respectively) are fixed, there is an optimal opening strength at which the coherent transport enhancement is optimized. In Secs.~\ref{sec-chain} and \ref{sec-fc}, respectively,
we extend our analysis to two paradigmatic models of transport: the linear chain and the fully connected
network. The linear chain in particular has been widely considered in the literature~\cite{cao1d,wusilbeycao,deph,plenio2}, and the fully connected network has been explored in Ref.~\cite{plenio2}. Finally, in Sec.~\ref{sec-fmo} we consider the Fenna-Matthews-Olson (FMO) light-harvesting complex, and demonstrate that the opening-assisted coherent transport obtained analytically in the earlier models is also present in this naturally occurring system. In Sec.~\ref{sec:6} we present our conclusions.

\section{Quantum networks}\label{sec:2}  
\begin{figure}[t]
\vspace{0cm}
\includegraphics[width=8cm,angle=0]{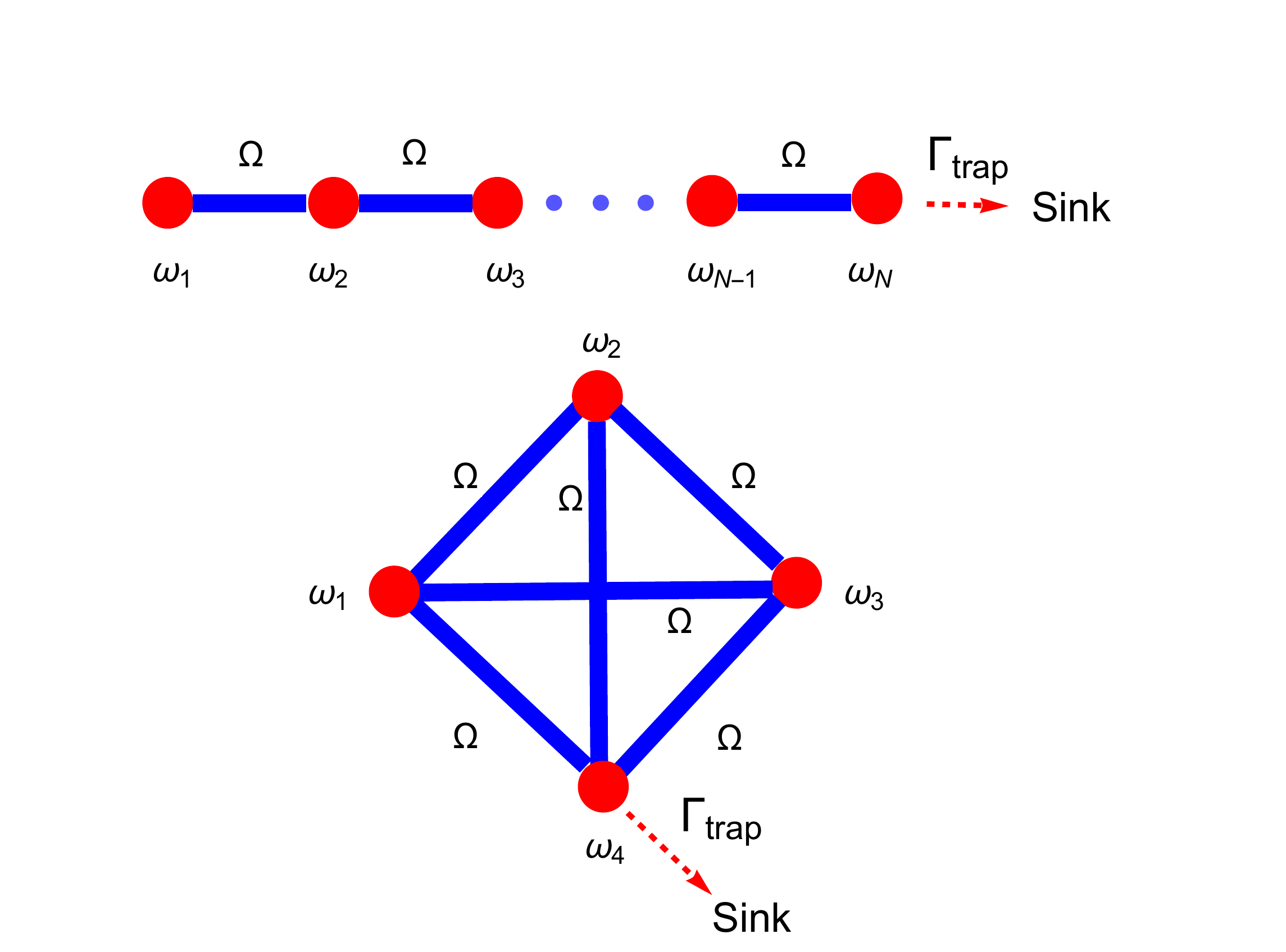}
\caption{(Color online)
Upper panel: The linear chain. One excitation
can hop between $N$ sites, with on-site energies $\omega_i$ and with
nearest neighbors 
connected by tunneling transition amplitude  $\Omega$.
Site $N$ is connected to a
decay channel,
where the excitation can escape,
 with coupling strength $\Gamma_{\rm trap}$. Lower panel: The analogous fully connected model with $N=4$ sites.
}
\label{network}
\end{figure}

Here we present the  quantum network models that we will
consider. A quantum network is a tight binding model where an
excitation can hop from site to site in a specified geometry.

The first example is the linear chain, see
the upper panel of Fig.~\ref{network}. 
This model has been widely analyzed in the literature due to its
relevance in natural and artificial energy transport devices,
and is characterized by the following
system  Hamiltonian ($\hbar=1$ here and in the following):
\begin{equation}
\text{H}_{\text{lin}}=\sum_{i=1}^{N}  \omega_{i} \ket{i}\bra{i}+
\Omega \sum_{j=1}^{N-1}  \left( \ket{j}\bra{j+1}+ \ket{j+1}\bra{j} \right)\,,  
\label{linham}
\end{equation}
where $\omega_{i}$ are the site energies and $\Omega$ is the coupling 
between neighboring sites. 
Here,  $|j\rangle$ represents  a state in which the excitation is at the site $j$,
when all the other sites are unoccupied. In terms of two-level states, $|0\rangle, 
|1\rangle$, it can be written  as
$ |j\rangle = |0\rangle_1 |0\rangle_2 \ldots |1\rangle_j \ldots |0\rangle_N.$
It is common to introduce static noise by letting
the energies $\omega_i$ fluctuate randomly in the interval $[-W/2, W/2]$ with a uniform distribution,
and variance $\sigma^2 = W^2/12$. 

This model can be ``opened'' by allowing
the excitation  to escape the system from one or more sites 
into continuum
channels describing 
the reaction center where the excitation is lost. 
This situation of ``coherent dissipation'' is applicable to
many systems and has been recently considered in~Ref.~\cite{alberto},
where it has been shown to give rise to the following 
effective non-Hermitian Hamiltonian (see also~\cite{rotterb}): 
\begin{equation}
(H_{\mathrm{eff}})_{jk}=(\text{H}_{\text{sys}})_{jk} -\frac{i}{2}  \sum_c A_j^c (A_k^c)^* \equiv
 (\text{H}_{\text{sys}})_{jk} -\frac{i}{ 2} Q_{jk}\,,
\label{amef}
\end{equation}
where $\text{H}_{\text{sys}}$ is the closed system Hamiltonian, e.g. $\text{H}_{\text{sys}}=\text{H}_{\text{lin}}$, and
$A_i^c$ are the transition amplitudes
from the discrete states $i$ to the continuum channels $c$.
If we consider a single decay channel, $c=1$, coupled to site $N$ with decay rate $\Gamma_{\text{trap}}$,
we have $A_N^1= \sqrt{\Gamma_{\text{trap}}/ 2}$,  and $Q_{jk}= \Gamma_{\text{trap}} \delta_{jN} \delta_{kN}$. Including fluorescence
effects, where the excitation may be lost from any site with rate $\Gamma_{\text{fl}}$, we have
$Q_{jk}= \left( \Gamma_{\text{trap}} \delta_{jN} +\Gamma_{\text{fl}}\right)\delta_{jk}$.

The quantum evolution (given by the operator ${\cal U } = e^{-iH_{\mathrm{eff}}t}$)
is non-unitary, since there is a loss of probability due to the decay channel and fluorescence.
The complex eigenvalues of $H_{\rm eff}$ can be written as
$E_r -i \Gamma_r/2$, 
where  $\Gamma_r$ represent the decay widths of the resonances. 
Superradiance, as discussed in the literature~\cite{puebla,Zannals}, 
 is usually reached only above a critical coupling strength with the continuum (in the
 overlapping resonance regime):
\begin{equation}
\label{orc}
\langle \Gamma \rangle/D \ge 1\,,
\end{equation}
 where $\langle \Gamma \rangle$ is the average decay width and $D$ is the
 mean level spacing of the closed system described by $\text{H}_{\text{sys}}$.

As a further effect of the environment 
we consider the dephasing caused by dynamic disorder. To include dephasing, we need to switch to  a master equation
for the reduced density matrix $\rho$~\cite{Haken-Strobl},
\begin{equation}\label{master}
	\dot{\rho}(t)=-\mathcal{L}_{\text{tot}} \rho(t),
\end{equation}
where the Liouville superoperator is given by $\mathcal{L}_{\text{tot}}=\mathcal{L}_{\text{sys}}+\mathcal{L}_{\text{trap}}+\mathcal{L}_{\text{fl}}+\mathcal{L}_{\text{deph}}$
and the four terms respectively describe the dynamics of the closed system, 
\begin{equation}
\mathcal{L}_{\text{sys}}\rho=i\left[\text{H}_{\text{sys}},\rho \right]\,,
\end{equation}
exciton trapping to the reaction center,
\begin{equation}
\mathcal{L}_{\text{trap}}\rho=\frac{\Gamma_\text{trap}}{2} \left\lbrace \ket{N}\bra{N},\rho \right\rbrace\,,
\end{equation}
decay due to fluorescence,
\begin{equation}
\mathcal{L}_{\text{fl}}\rho=\Gamma_\text{fl} \rho\,,
\end{equation}
and the dephasing effect as described in the simplest approximation by the Haken-Strobl-Reineker (HSR) model~\cite{Haken-Strobl} with dephasing rate $\gamma$,
\begin{equation}
(\mathcal{L}_{\text{deph}}\rho)_{jk} = \gamma \rho_{jk} \left(1-\delta_{jk}\right) \,.
\end{equation}  

The efficiency of exciton transport can be measured by the total population trapped by the sink~\cite{deph,plenio2},
\begin{equation}
\eta=\Gamma_{\text{trap}} \int_{0}^{\infty} \rho_{NN}(t) \,dt  \,,
\label{eqeta}
\end{equation}
or by the average transfer time to reach the sink~\cite{lloyd},
\begin{equation}
\tau=\frac{\Gamma_{\text{trap}}}{\eta} \int_{0}^{\infty} t\,
\rho_{NN}(t) \,dt\,.
\label{eqtau}
\end{equation}

The system is initiated with one exciton at site $1$, i.e.,  $\rho(0)=\ket{1}\bra{1}$. 
Formally  the solutions for $\eta$ and $\tau$ can be written as,
\begin{equation}
\eta=\Gamma_{\text{trap}}(\mathcal{L}_{\text{tot}}^{-1}
\rho(0))_{\text{NN}}
\end{equation}
and 
\begin{equation}
\tau=\frac{\Gamma_{\text{trap}}}{\eta}(\mathcal{L}_{\text{tot}}^{-2} \rho(0))_{\text{NN}}\,.
\label{tau_liu}
\end{equation}
In physical applications, we are typically interested in the parameter regime of high efficiency $\eta$, which can occur only when fluorescence is weak, i.e., when the fluorescence rate $\Gamma_{\text{fl}}$ is smaller than both the trapping rate $\Gamma_{\text{trap}}$ and the energy scales in the closed-system Hamiltonian $\text{H}_{\text{sys}}$. The FMO complex discussed in Sec.~\ref{sec-fmo} is a typical example: here the exciton recombination time $1/\Gamma_{\text{fl}}$ is estimated to be around 1 ns, whereas the other times scales in the problem are of the order of picoseconds or tens of picoseconds~\cite{lloyd,deph,plenio2}. In this
regime, the effect of $\Gamma_{\text{fl}}$ on the efficiency $\eta$ and transfer time $\tau$ may be treated perturbatively (see e.g. Ref.~\cite{caosilbey}): Specifically, $\tau$ is independent of $\Gamma_{\text{fl}}$ to leading order, and $\eta$ is related to $\tau$ by
\begin{equation}
\label{eta_tau}
\eta = \frac{1}{1+\Gamma_{\text{fl}} \tau}
\end{equation}
when higher-order corrections are omitted. Thus, for a given fluorescence rate, maximizing efficiency $\eta$ is entirely equivalent to minimizing the transfer time $\tau$. In the following, we will assume for simplicity of presentation that $\Gamma_{\text{fl}}$ is indeed small, and will present results for $\tau$ only; analogous expressions for the efficiency $\eta$ may be easily obtained by inserting these results into Eq.~(\ref{eta_tau}).

In the following, we will be interested in the disorder-ensemble averaged transfer time, defined as
\begin{equation}
\left<\tau\right>_W= \frac{1}{W^N}\int_{-W/2}^{W/2}..\int_{-W/2}^{W/2}
\tau(\omega_1,\omega_2,..\omega_N) \, d\omega_1d\omega_2 \ldots d\omega_N\,.
\label{tauW}
\end{equation}   

\section{Two-site model}
\label{sec-2site}

\subsection{F\"orster approximation}

In the 1940s, F\"orster~\cite{forster} proposed  an incoherent non-radiative
resonance theory of the energy transfer process in weakly coupled
pigments.  This mechanism was based on the assumption that, due to
large dephasing, the motion of an excitation between
chromophores is a classical random walk, which can be described by an incoherent master equation.

Let us first consider a dimer of interacting chromophores and the transmission of the excitation from one molecule to the other. 
The Hamiltonian of the system is
\begin{equation}H=  \begin{pmatrix} \omega_1    &   \Omega \\   \Omega   &    \omega_2 \end{pmatrix} \,,\end{equation}
where $\Omega$ and $ \omega_1-\omega_2=\Delta$ are respectively the coupling and the excitation energy 
difference between the two molecules. Note that $|1\rangle$ represents
a state where molecule $1$ is excited and molecule $2$ is in its ground state.

The  energy difference or detuning $\Delta$ is entirely due to the interaction with the environment, if we assume the molecules of the dimer to be identical.
The exciton-coupled dimer is most productively viewed as a
supermolecule with two delocalized electronic transitions, rather than
a pair of individual molecules, which means switching to the basis
that diagonalizes $H$.

For this Hamiltonian, the probability for an initial 
excitation in the first molecule to move to the second one is given by
\begin{equation}
P_{1 \rightarrow 2} (t) = \frac{4 \Omega^2}{4 \Omega^2+\Delta^2}
\sin^2  \left( \sqrt{4 \Omega^2+\Delta^2}  t/2 \right) \,,
\end{equation}
to which we can associate a typical hopping time 
$
\tau_{\rm hop} = \pi/\sqrt{4 \Omega^2+\Delta^2}
$,
a very important parameter for understanding the propagation.

In the F\"orster theory, dephasing is assumed to be large. If $\gamma \gg 1/\tau_{\rm hop}$, the dephasing time is much smaller than the hopping time, $\tau_d= 1/\gamma \ll \tau_{\rm hop}$. In this regime, coherence is suppressed and exciton dynamics becomes diffusive. The transfer rate from one molecule to the other is given by:
\begin{equation}
T_{1\to2} \sim \frac{d P_{1\to2}(\tau_d)}{ d \tau_d}  \approx\frac{2 \Omega^2}{\gamma}  \,.
\label{T1}
\end{equation}
This transfer rate also gives the diffusion coefficient for a linear chain of chromophores coupled by a nearest-neighbor interaction, as considered in Refs.~\cite{4goldi,cao1d} for $\Omega \gg \Delta$. Indeed, the mean squared number of steps that an excitation can move is proportional to the time measured in units of the average transfer time $\tau= 1/T_{1\to2}$, i.e., $r^2(t) \propto t/\tau= T_{1\to2} t$. The diffusion coefficient in this regime is thus given by Eq.~(\ref{T1}) and it agrees with previous results~\cite{4goldi,cao1d} in the same regime.

If dephasing is still large compared to the coupling $\Omega$, but small compared to the detuning $\Delta$,  $\Delta \gg \gamma \gg \Omega$ , we must average $P_{1\to2}(t)$ over time and obtain
\begin{equation}
T_{1\to2}= \frac{\overline{P_{1\to2}}}{\tau_d} \approx\frac{2 \Omega^2 \gamma}{\Delta^2}  
\label{T2}
\end{equation}
This expression also agrees with the diffusion coefficient given in~\cite{4goldi,cao1d} in the same regime. 

In general, as long as  $\gamma \gg \Omega$ holds, we have the F\"orster transition rate
\begin{equation}
T_{\rm F} = \frac{2 \Omega^2 \gamma}{
	\gamma^2 +\Delta^2} \,,
\label{Lcl}
\end{equation}
with the scalings given by Eqs.~(\ref{T1}) and (\ref{T2}) as special cases.

Here we will not discuss the weak dephasing regime $\gamma<\Omega$ in which F\"orster theory does not apply. This regime has been investigated in~\cite{4goldi,cao1d}, where it was shown that  the excitation dynamics is still diffusive, but with mean free path of order the localization length, so that the diffusion coefficient is enhanced by the localization length squared.

\subsection{Two-site model with opening}
\begin{figure}
\includegraphics[width=11cm]{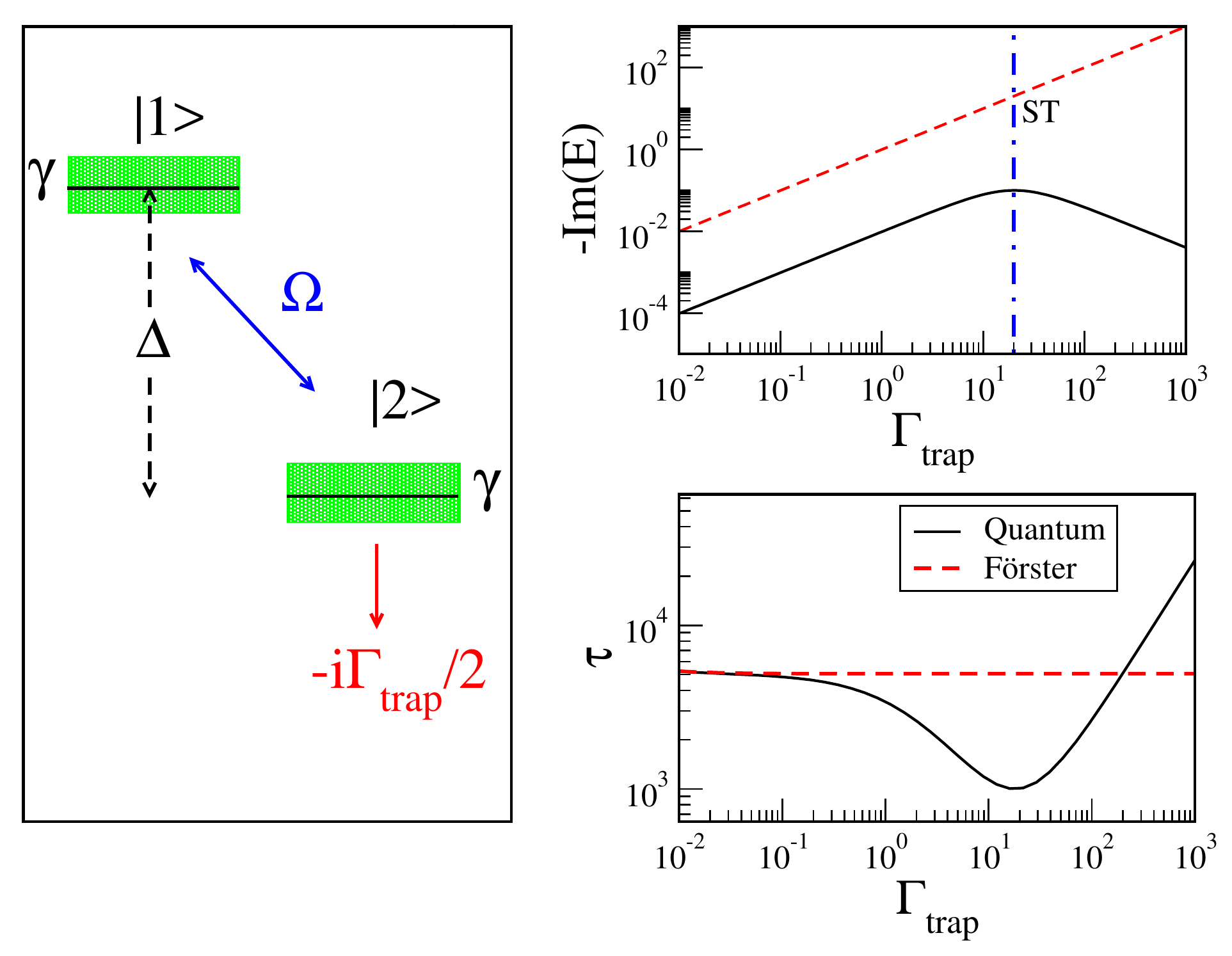}
\caption{(Color online) Left panel: Schematic view of the two-site model in the presence
  of dephasing and coupling to the sink. Upper right panel: The imaginary
  part of the eigenvalues of the non-Hermitian Hamiltonian given in
  Eq.~(\ref{E1}). Here $\Omega=1,
  \Delta=10, \gamma=0$. Lower right panel: Transfer time as a function of
  decay width to the sink for the quantum model (solid curve) and for
  the F\"orster model (dashed curve). The vertical dashed line represents the superradiance transition (ST). Here $\Omega=0.1$, $\gamma=1,
  \Delta=10$. 
}
\label{fig2x2}
\end{figure}

The same two-site system can be considered in the most general context
in which the interaction with a sink is explicitly taken into
account. 
For this purpose we add
to the two-site Hamiltonian described in the previous section 
a term representing the possibility of escaping from state $|2\rangle$ to an external continuum
 with decay rate $\Gamma_{\rm trap}$, see Fig.~\ref{fig2x2} (left panel).
 Moreover
the system is in contact with another environment that induces fast time-dependent
fluctuations of the site energies with variance proportional to
$\gamma$. The presence of detuning strongly suppresses the
probability of the excitation leaving the system. On the other hand, dephasing produces an energy broadening,
which facilitates transport. For very large dephasing, the probability
for the two site energies to match becomes small and thus
transport is again suppressed. Optimal transport thus occurs at some
intermediate dephasing value:  $\gamma \approx \Delta$~\cite{lloyd,deph}. This
is the noise assisted transport: Noise can help in a situation where transport
is suppressed in presence of only coherent motion. 

Another general principle that is essential for understanding transport efficiency in open systems is
superradiance. Indeed, due to the coupling with a continuum of states,
the state $|2\rangle$ has an energy broadening $\Gamma_{\rm trap}$, even in absence of dephasing,
which can also facilitate transport. The system in the absence of dephasing is
described by the following $2 \times 2$ effective non-Hermitian
Hamiltonian: 
\begin{equation}
H_{\rm eff} = \left( \begin{array}{cc}
\omega_1 & \Omega\\
\Omega & \omega_2 -i \Gamma_{\rm trap}/2 \\
\end{array} \right) \,.
\label{E1}
\end{equation}

The complex eigenvalues (taking $\omega_1=0$ and $\omega_2=\Delta$) are:
\begin{equation}
{\cal E}_{\pm}= \frac{\Delta}{2}-i \frac{\Gamma_{\rm trap}}{4} \pm
\frac{1}{2} \sqrt{(\Delta-i\frac{\Gamma_{\rm trap}}{2})^2+4 \Omega^2} \,,
\end{equation}
and their  imaginary parts 
represent the decay widths of the system. As a function of
$\Gamma_{\rm trap}$, one of the decay widths has a non-monotonic
behavior which signals the superradiance transition (ST), see
Fig.~\ref{fig2x2} (upper right panel).  For $\Delta \gg
\Omega$, this transition, corresponding to the maximum of the smaller width,
occurs at $\Gamma_{\rm trap} \approx 2 \Delta$, see the dashed vertical
line in  Fig.~\ref{fig2x2} (upper right panel).
Transport efficiency is optimized around the ST, where the transfer time has a minimum as shown in Fig.~\ref{fig2x2} (lower
right panel). Indeed for small $\Gamma_{\rm trap}$, the transport becomes more efficient
with increasing $\Gamma_{\rm trap}$, since the decay width of both
states increases. On the other side, above the ST, only one of the two decay
widths continues to increase with $\Gamma_{\rm trap}$, while the
other decreases. At the same time, the state with the larger decay
width becomes localized on site $|2\rangle$, thus suppressing
transport. 
 
Note that while noise-assisted transport occurs only in presence of
a detuning $\Delta$, superradiance-assisted transport (SAT) occurs even with
$\Delta=0$ and in the absence of dephasing.

The non-monotonic behavior of the transfer time as a function of
$\Gamma_{\rm trap}$ is a purely quantum coherent effect. To see this effect analytically, we consider the master equation (\ref{master}), which for the two-site model can be written explicitly as
\begin{equation}
\left( 
\begin{array}{c} 
\dot{\rho}_{11} \\  \dot{\rho}_{12} \\ \dot{\rho}_{21} \\ \dot{\rho}_{22} 
\end{array} 
\right)=
\left( \begin{array}{cccc}
0& i\Omega & -i \Omega&0\\
i\Omega & i\Delta - \frac{\Gamma_{\rm trap}}{2} -\gamma& 0& -i\Omega \\
-i \Omega &0 & -i\Delta  - \frac{\Gamma_{\rm trap}}{2} -\gamma & i \Omega \\
0&-i\Omega& i\Omega& -\Gamma_{\rm trap}
\end{array} \right)
\left ( \begin{array}{c} \rho_{11} \\ \rho_{12} \\
   \rho_{21} \\ \rho_{22} \end{array} \right) \,.
\label{M2}
\end{equation}
Following \cite{caosilbey} we may insert the stationary solution
($\dot{\rho}_{12}=\dot{\rho}_{21}=0$) for the
off-diagonal matrix elements into Eq.~(\ref{M2}) and obtain a rate
equation for the populations ${\rho}_{11}$ and ${\rho}_{22}$ only: 
\begin{equation}
\left( 
\begin{array}{c} 
\dot{\rho}_{11} \\  \dot{\rho}_{22} 
\end{array} 
\right)=
\left( \begin{array}{cc}
-T_{1\to2}& T_{2\to1} \\
T_{1\to2} & -T_{2\to1} -\Gamma_{\rm trap}
\end{array} \right)
\left ( \begin{array}{c} \rho_{11} \\ \rho_{22} \end{array} \right) \,.
\label{M22}
\end{equation}
These transition rates have been derived by Leegwater in \cite{leegwater}. 
In our case we have $T_{1\to2}=T_{2\to1}=T_{\rm L}$ with 
\begin{equation}
T_{\rm L} = \frac{2 \Omega^2 (\gamma+\Gamma_{\rm trap}/2)}{
  (\gamma+\Gamma_{\rm trap}/2)^2 +\Delta^2} \,.
\label{Lqu}
\end{equation}
The incoherent master equation given in Eq.~(\ref{M22}) represents  a good
approximation of the exact quantum dynamics, Eq.~(\ref{M2}), when the off-diagonal matrix
elements reach a stationary solution very fast. 
This is valid when the dephasing is sufficiently fast:
\begin{equation}
\gamma \gg \Omega \,,
\label{F1}
\end{equation}
which is the same condition as the one that ensures validity of the F\"orster transition rate approximation (\ref{Lcl}) in the closed system. We observe that the Leegwater rate given by Eq.~(\ref{F1}) reduces to the F\"orster rate given by Eq.~(\ref{Lcl}) in the limit where the system is closed, $\Gamma_{\rm trap} \to 0$.

Here we would like to stress an important point: in a classical model
of diffusion the transition rates from site to site are completely
independent of the escape rates associated with individual sites. 
For this reason the transition rates given in
Eq.~(\ref{Lqu}) cannot  correspond to any classical
diffusion model due to their dependence on $\Gamma_{\rm trap}$. 
So the transition probability given in 
Eq.~(\ref{Lqu}) includes also coherent effects due to the opening. 
This point of view is slightly different from the one in
\cite{caosilbey} where the master equation
Eq.~(\ref{M22}) is viewed as ``classical.'' From now on we will refer to the master equation (\ref{M2})  with transition rates given
in Eq.~(\ref{Lqu}) as the Leegwater
model, while the  F\"orster model will denote the master equation (\ref{M2}) with the F\"orster transition
rates (\ref{Lcl}), independent of $\Gamma_{\rm trap}$. Needless to say, $T_{\rm L}(\Gamma_{\rm trap}=0)=T_{\rm F}$.

Now two questions present themselves. First, we would like to understand which values of the opening strength $\Gamma_{\rm
  trap}$ cause the F\"orster model to fail due to the coherent effects
induced by the opening. Comparing Eqs.~(\ref{Lcl}) and (\ref{Lqu}), it is clear that the F\"orster
model applies when Eq.~(\ref{F1}) holds and $\Gamma_{\rm trap}/2 \ll
\gamma$. Even in the presence of large dephasing, when $\Gamma_{\rm trap}$ is also large (and
becomes of the order of $\gamma$), coherent effects cannot be neglected
and quantum transport differs significantly from that predicted by the F\"orster theory (compare
the red dashed curve with the solid black curve in  Fig.~\ref{fig2x2} (lower
right panel).

Second, we would like to address whether quantum effects
can provide enhancement over the transport predicted by F\"orster theory. A clear example
showing that this can happen appears in Fig.~\ref{fig2x2} (lower right
panel), where, for a large region of values of $\Gamma_{\rm trap}$,
the quantum transfer time is significantly less than that predicted by F\"orster
theory. 
So the idea is the following: Even in presence of large dephasing, for
which a F\"orster model of incoherent transport is expected to
apply, as we increase the coupling $\Gamma_{\rm trap}$ to a sink,
coherent effects can be revived and enhance transport. 
Finding overall conditions for optimal transport in open quantum systems will be a key focus of the following analysis.

Below we will derive analytical expressions for the
transfer times and address the above questions quantitatively.

\subsection{Transfer time, optimal opening, and quantum enhancement}

In the two-site case, one can obtain a simple yet exact analytic form for
the transport time $\tau$, Eq.~(\ref{eqtau}), using Eq.~(\ref{tau_liu}) and substituting the exact Liouville operator $\mathcal{L}$ given by 
Eq.~(\ref{M2})~\cite{caosilbey}:

\begin{equation}\label{tau}
\tau=\frac{1}{2\Omega^2}
\left( \frac{4\Omega^2}{\Gamma_{\text{trap}}}+\gamma+\frac{\Gamma_{\text{trap}}}{2}+\frac{\Delta^2}{\gamma+ \frac{\Gamma_{\text{trap}}}{2}}\right)\,.
\end{equation}
Eq.~(\ref{tau}) shows explicitly the non-monotonic behavior of the transfer time with the opening $\Gamma_{\text{trap}}$, which is a signature of quantum coherence and is clearly visible in Fig.~\ref{fig2x2} (lower right panel).

The expression for the average transfer time can be aso computed using the
incoherent master equation (\ref{M2}), with either the F\"orster or
Leegwater transition rate. While for the two-site case the Leegwater
average transfer time is exactly the same as the full quantum
result (\ref{tau}), for the F\"orster theory we have: 
\begin{equation}\label{tau_cl}
\tau_{\rm F}=\frac{1}{2\Omega^2}
\left( \frac{4\Omega^2}{\Gamma_{\text{trap}}}+\gamma+\frac{\Delta^2}{\gamma}\right)\,.
\end{equation}
We note that $\tau_{\rm F}$ decays monotonically with increasing opening $\Gamma_{\text{trap}}$, as it must in a classical calculation. Clearly for $\Gamma_{\rm trap} \ll \gamma$, F\"orster theory coincides
with the full quantum result. On the other hand for $\Gamma_{\rm trap}
\gtrsim \gamma$, coherent effects become important and they can be
incorporated using the Leegwater model (at least for the two-site case). 

\begin{figure}
\includegraphics[width=12cm]{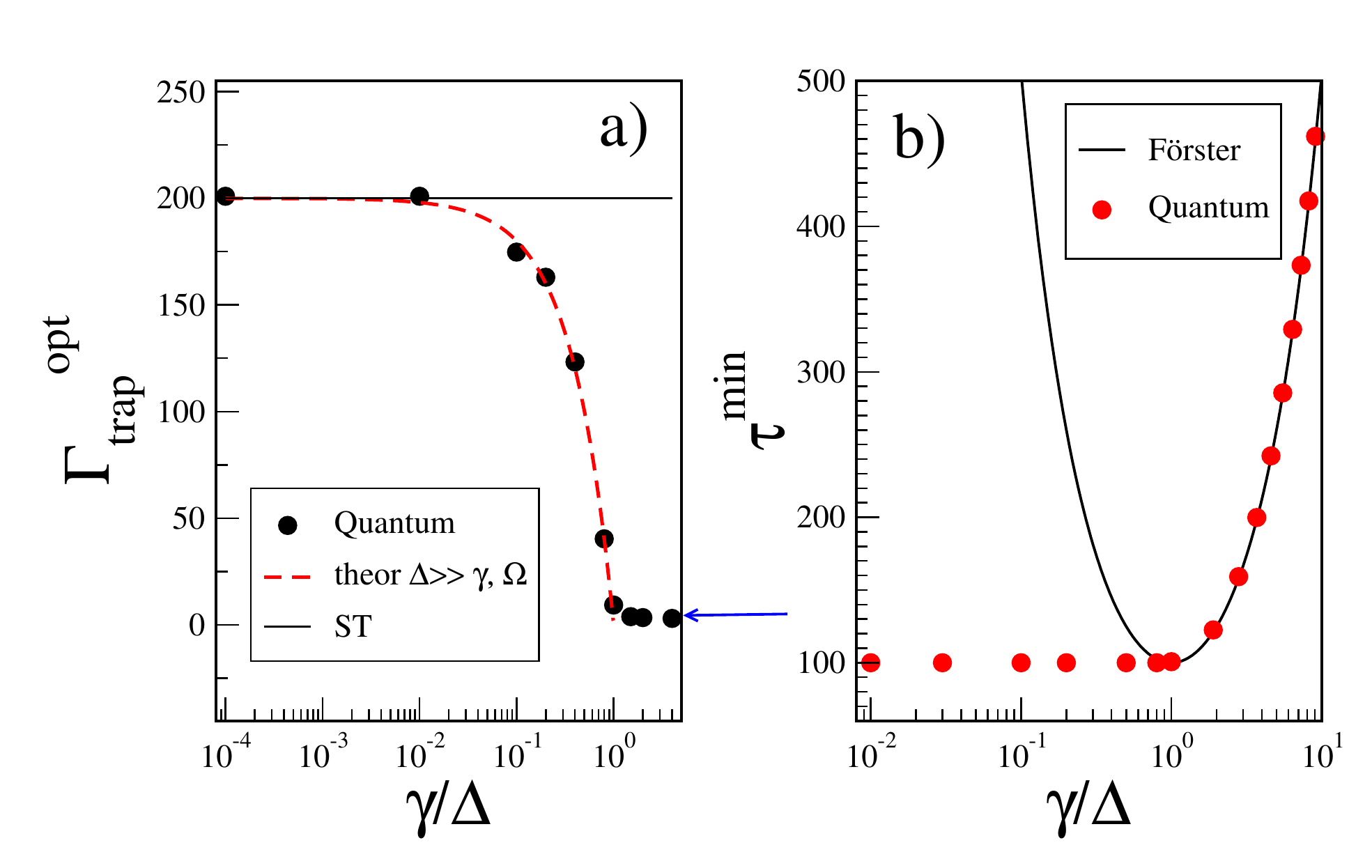}
\caption{(Color online) Left panel:  Optimal coupling to the sink, $\Gamma_{\rm trap}^{\rm opt}$, in a two-site system, as a
  function of the rescaled dephasing strength $\gamma/\Delta$ in the regime $\Delta
  \gg \Omega$. Data refer to the case $\Delta=100, \Omega=1$. Symbols
  represent numerical simulations of the full quantum model, the dashed
  red curve shows the analytical result given by Eq.~(\ref{Gopt}), and the
  blue arrow shows the asymptotic value given by Eq.~(\ref{as}). The solid horizontal line
  indicates  the value at which the superradiance transition (ST) occurs for
  zero dephasing. Right panel: Minimal transfer times for the
  F\"orster model (solid curve) and for the full quantum calculation (symbols)
  are shown as functions of the rescaled dephasing strength $\gamma/\Delta$.
}
\label{f1opt}
\end{figure}

Since $\tau_{\rm F}$ is a monotonic function 
of $\Gamma_{\rm trap}$, it assumes its minimum value 
\begin{equation}
\tau_{\rm F}^{\rm min}=\frac{1}{2\Omega^2} \left( \gamma+\frac{\Delta^2}{\gamma}\right)
\label{tF}
\end{equation}
for $\Gamma_{\rm trap}\to\infty$.
On the other hand,  the quantum transfer time is minimized at a finite value of
$\Gamma_{\rm trap}$. Unfortunately the optimal value of $\Gamma_{\rm trap}$ is given in general by the solution to a quartic equation. Nevertheless it is easy to obtain simple
expressions in several physically relevant regimes. In particular, of greatest physical interest is the situation where the quantum minimum associated with optimal value of the opening is deep, which is only possible where a large difference exists in the first place between quantum and incoherent transport, i.e.,
$\Gamma_{\rm trap} \gg \gamma$, as discussed above in Sec.~\ref{sec-2site}. In that regime, the optimal opening is given by
\begin{equation}
\Gamma_{\rm trap}^{\rm opt} \approx 2 \sqrt{\Delta^2+2 \Omega^2} -\frac{2\gamma \Delta^2}{\Delta^2+2 \Omega^2} +O(\gamma^2) \,.
\label{Goptsmg}
\end{equation}
If in addition to dephasing being weak, detuning is strong ($\Delta \gg \Omega$), Eq.~(\ref{Goptsmg}) simplifies to
\begin{equation}
\Gamma_{\rm trap}^{\rm opt} \approx 2 \Delta -2\gamma \,.
\label{Gopt}
\end{equation}
In Fig.~\ref{f1opt} (left panel) the simple analytical expression (\ref{Gopt}) is shown to agree
very well with exact numerical calculations for the quantum model. This result is
particularly interesting since it shows the effect of dephasing on the
ST: While for small dephasing the optimal opening strength is given by
the ST criterion $\Gamma_{\rm ST} \approx 2 \Delta$, for larger dephasing the optimal $\Gamma_{\rm
  trap}=\Gamma_{\rm ST} -2 \gamma$ decreases with the dephasing $\gamma$. 

The condition for optimal transport given in  Eq.~(\ref{Gopt}), can be
re-written as $\Delta= \gamma + \Gamma_{\rm trap}^{\rm opt}/2$. This can be
interpreted by saying that dephasing and opening together
induce a cumulative energy broadening, which optimize
transport when it matches the detuning $\Delta$.  Also striking is the
symmetrical role that $\gamma$ and $\Gamma_{\rm trap}$ play in controlling transport efficiency even if
their origin and underlying physics are completely different. For instance $\gamma$ induces
dephasing in the system, whereas $\Gamma_{\rm trap}$ increases the
coherent effects.

The optimal dephasing, fixing all other variables, is given exactly by
\begin{equation}
\gamma^{\rm opt}= \Delta- \Gamma_{\rm trap}/2 \,,
\label{gopt1}
\end{equation}
 in any regime. This shows that also the criterion
for noise assisted transport, $\gamma \approx \Delta$, is modified by the
presence of a strong opening. 

For the value of $\Gamma_{\rm trap}$ given in Eq.~(\ref{Gopt}),  the minimal transfer time assumes the value: 
\begin{equation}
\tau^{\rm min} \approx \frac{\Delta}{\Omega^2} \,,
\end{equation}
which should be compared with the F\"orster expression (\ref{tF}) in
the same regime,
\begin{equation}
\tau_{\rm F}^{\rm min} \approx \frac{\Delta^2}{2 \gamma \Omega^2} = \tau^{\rm min}
\frac{\Delta}{2 \gamma} \gg \tau^{\rm min} \,,
\end{equation}
showing that in this regime ($\Delta \gg \gamma \gg \Omega$) quantum coherence, induced by the
coupling to the sink, can always enhance transport. 

In the opposite limit $\Omega \gg \Delta \gg \gamma$  one obtains: 
\begin{equation}
\Gamma_{\rm trap}^{\rm opt}= 2 \sqrt{2} \Omega \,.
\label{as}
\end{equation}

We summarize our  results so far in the following way: For very small opening,
$\Gamma_{\rm trap} \ll \gamma$, the F\"orster model  reproduces the quantum
results. In the opposite limit $\Gamma_{\rm trap}\to\infty$, quantum transport is always fully
suppressed while the F\"orster model prediction for the average transfer time approaches a non-zero asymptotic value,
thus showing the non-applicability of this model.  In general, we expect the F\"orster
model to fail when $\Gamma_{\rm trap} \gtrsim \gamma$, so that
coherent effects that occur on a time scale $1/\Gamma_{\rm trap}$,
can be relevant before dephasing destroys them on a time scale
$1/\gamma$.

What is the regime in which quantum 
transport is better than the incoherent transport described by the F\"orster model? In order to find this regime,
let us write the difference between the two transfer times: 
\begin{equation}\label{diff}
	\tau_{\rm F}-\tau=\frac{\Gamma_{\text{trap}}}{4\Omega^2}
        \left[
          \frac{\Delta^2}{\gamma(\gamma+\Gamma_{\text{trap}}/2)}-1 \right] \,.
\end{equation}

\begin{figure}
	\includegraphics[width=12cm]{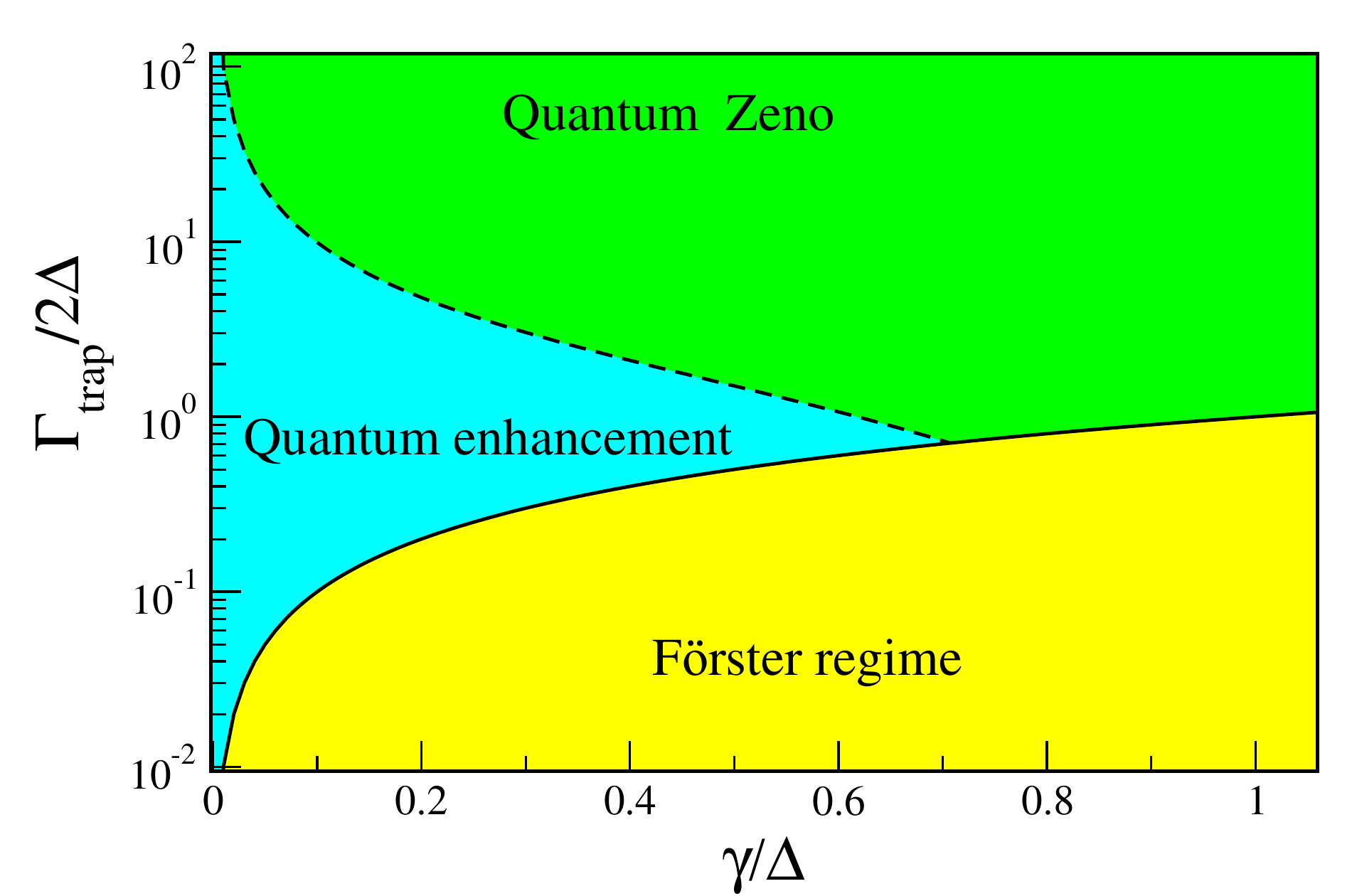}
	\caption{(Color online) The parameter regime of significant quantum enhancement of transport in the two-site model, Eq.~(\ref{g22}), and the F\"orster regime, $\gamma < \Gamma_{\text{trap}}/2$, where quantum effects are negligible. In the third regime, corresponding to very large opening $\Gamma_{\text{trap}}$, quantum mechanics suppresses transport due to the quantum Zeno effect. Generalization to a disordered linear chain of arbitrary length is obtained by replacing the detuning $\Delta$ in the two-site model with $W/\sqrt{6}$, where $W$ is the disorder strength.
	}
	\label{figphase}
\end{figure}

Clearly, quantum transfer is enhanced over the F\"orster prediction ($\tau < \tau_{\rm F}$) if and only if 
$\gamma+\Gamma_{\text{trap}}/2 < \Delta^2/\gamma$. On the other hand, as noted earlier, the relative difference is small, i.e., the F\"orster model is a good approximation, when $\Gamma_{\rm trap}/2 \ll \gamma$. So the regime where quantum coherent effects produce a significant enhancement of
transfer efficiency in the two-site model is given by
\begin{equation}
\gamma \lesssim \frac{\Gamma_{\text{trap}}}{2} < \frac{\Delta^2}{\gamma}-\gamma \,.
\label{g22}
\end{equation}
This result is consistent with the illustration in Fig.~\ref{fig2x2} (lower right panel): when the opening is very small, the F\"orster approximation holds, whereas for very large opening,  coherent effects cause trapping (a quantum Zeno effect). It is only for the range of openings given in Eq.~(\ref{g22}) that quantum coherence aids transport.

The quantum transport enhancement regime given by Eq.(~\ref{g22}) is illustrated graphically in Fig.~\ref{figphase}. As the dephasing decreases, quantum enhancement of transport occurs for an ever wider range of openings $\Gamma_{\text{trap}}$. We also observe that near the superradiance transition, $\Gamma_{\text{trap}} \sim 2 \Delta$, quantum transport enhancement obtains for the widest range of dephasing strengths $\gamma$.

Note that in the case of static disorder,  the disorder-averaged transfer time can be
computed as stated in Eq.~(\ref{tauW}). The results of this section
remain valid if we    substitute $\Delta^2$ with  $\langle (\omega_1-\omega_2)^2 \rangle=W^2/6$.

\section{Long chains with static disorder}
\label{sec-chain}

\subsection{Linear chain: Analytic results}

For a linear chain with $N$ sites, see Fig.~\ref{network} (upper panel), in the presence of static
disorder, it is not possible to get an analytical expression for the
full quantum model. 
Nevertheless under the strong dephasing condition given in
Eq.~(\ref{F1})  the dynamics of the system can be described by
 the incoherent master equation,
\begin{equation}\label{P}
\frac{dP_j}{dt}=\sum_{k}(T_{k\to j}P_k-T_{j\to k}P_j) -\delta_{j,n} \Gamma_{\text{trap}} P_j \,,
\end{equation}
where $P_j$ is the probability to be at site $j$.
The nearest-neighbor transfer rates in Eq.~(\ref{P}), $T_{k\to j}$, are given by $T_{\rm F}$, Eq.~(\ref{Lcl}), with the exception of the transfer rate along the bond adjacent to the sink, $T_{n-1\to n}=T_{n\to n-1}$, which is given by the Leegwater expression $T_{\rm L}$, Eq.~(\ref{Lqu}). The F\"orster model is also given by Eq.~(\ref{P}) but with all the transfer rates given by $T_{\rm F}$, Eq.~(\ref{Lcl}).

Proceeding in the same way as for the case $N=2$, we obtain 
analytical expressions for the ensemble-averaged F\"orster and Leegwater transfer times:
\begin{equation}\label{cl}
\left<\tau_{\text{F}}\right>_W=\frac{N}{\Gamma _{\text{trap}}}+\frac{N\left(N-1\right)}{4 \Omega^2}\left(\gamma+\frac{W^2}{6\gamma}\right)
\end{equation}
and 
\begin{widetext}
\begin{equation}\label{leeg}
\left<\tau_{\text{L}}\right>_W=\frac{N}{\Gamma _{\text{trap}}}+\frac{N\left(N-1\right)}{4 \Omega^2}\left[\gamma +\frac{\Gamma_{\rm trap}}{N}+\frac{W^2}{6\gamma}\left(1-\frac{2 \Gamma_{\rm trap}}{N(2 \gamma+ \Gamma_{\rm trap})} \right)\right] \,.
\end{equation}
\end{widetext}

The effect of quantum coherence is given by the difference in transfer times,  
\begin{equation}\label{leeg_chain}
\left<\tau_{\text{F}}\right>_W-\left<\tau_{\text{L}}\right>_W=\frac{(N-1) \Gamma _{\text{trap}}}{4 \Omega ^2} \left(\frac{W^2}{3 \gamma  \left(2
	\gamma +\Gamma _{\text{trap}}\right)}-1\right).
\end{equation}
In general, increasing the ratio $W/\gamma$ (i.e., increasing the strength of static as compared to dynamical disorder) will make the difference in Eq.~(\ref{leeg_chain})
more positive, i.e., quantum transport becomes more favored relative to incoherent transport, just as has been seen already in the two-site case. To be precise, from Eq.~(\ref{leeg_chain}),
$$\frac{W^2}{3 \gamma  \left(2\gamma +\Gamma _{\text{trap}}\right)}>1 \Rightarrow W> \sqrt{6}\gamma $$
must hold in order to have 
 $\left<\tau_{\text{F}}\right>_W > \left<\tau_{\text{L}}\right>_W$, and the regime where quantum effects are both helpful and significant is then identical to the one identified in Eq.~(\ref{g22}) and Fig.~\ref{figphase} for the two-site model, with the simple replacement $\Delta^2 \to W^2/6$:
 \begin{equation}
 \gamma \lesssim \frac{\Gamma_{\text{trap}}}{2} < \frac{W^2}{6\gamma}-\gamma \,.
 \label{qenhchain}
 \end{equation}
 We notice that $W>\sqrt{6}\gamma$ is a necessary condition for significant quantum transport enhancement to occur, i.e., static disorder must be stronger than dynamic disorder.
Given $W>\sqrt{6} \gamma $, 
$\left<\tau_{\text{F}}\right>_W-\left<\tau_{L}\right>_W$ is maximized when $\Gamma_{\text{trap}}=\Gamma_{\text{trap}}^{\rm opt}$, where
\begin{equation}
\label{gopt}
\Gamma_{\text{trap}}^{\rm opt}=2 \left(W/\sqrt{6}-\gamma\right) \,.
\end{equation} 
To be  precise, we should note that the value of $\Gamma_{\text{trap}}$ that maximizes the transport enhancement $\left<\tau_{\text{F}}\right>_W-\left<\tau_{L}\right>_W$ is not exactly the same
as the value at which the quantum transport time $\left<\tau_{L}\right>_W$ is minimized, but in the limit of small $\Omega$ where the F\"orster approximation is meaningful, the difference is negligible.
In the limiting case $W \sim \Gamma_{\text{trap}}\gg\gamma\gg \Omega$ 
and $N\gg1$, $\left<\tau_{\text{F}}\right>_W \approx \left<\tau_{\text{L}}\right>_W \approx 
N^2 W^2/24\gamma\Omega^2$, so both types of transport are diffusive, while the difference in transfer times 
 $\left<\tau_{\text{F}}\right>_W-\left<\tau_{L}\right>_W$
is $N W^2/12\gamma\Omega^2$.
In relative terms,  quantum enhancement is therefore most important in short
chains, which is a case relevant in realistic photosynthetic complexes
where the number of chromophores is small, e.g. the FMO complex which will be the focus of  Sec.~\ref{sec-fmo}.

\subsection{Linear chain: Numerical results}

Eqs. (\ref{cl}) and (\ref{leeg}) provide, respectively,
the analytical results for the
average transfer time in the F\"orster model, which is purely incoherent, and in the Leegwater
approximation, which incorporates quantum coherence effects. Unfortunately no analytic result is available for the
exact quantum calculation in a chain of general length $N$ and for this reason we will present results obtained by means of numerical simulations. In particular we will show that:
\begin{itemize}
\item The Leegwater expression, Eq.~(\ref{leeg}), provides a good approximation 
for the exact quantum transfer time in the regime of interest given by Eq.~(\ref{qenhchain}), as long as
the semiclassical condition, $\gamma \gg \Omega$, holds;
\item Equation~(\ref{gopt}), obtained from the analytic Leegwater calculation, accurately describes the opening at which the exact quantum transport enhancement $\langle \tau_{\text{F}} \rangle_W
- \langle \tau \rangle_W$ is maximized;
\item Quantum corrections beyond the Leegwater approximation  give rise to even
stronger quantum transport enhancement when the semiclassical condition $\gamma \gg \Omega$ is no longer
satisfied.
\end{itemize}

\begin{figure}
\includegraphics[width=11cm]{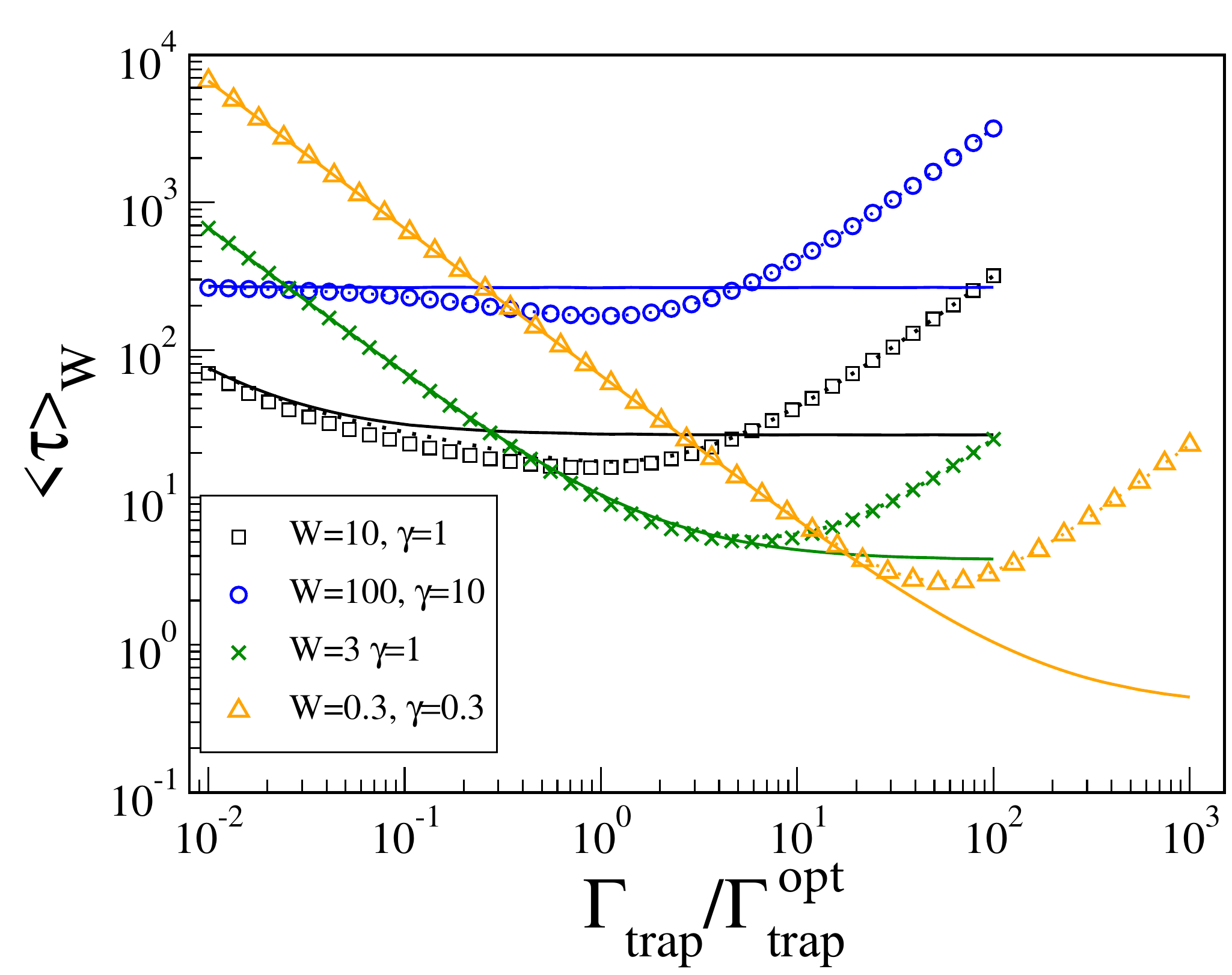}
\caption{(Color online) F\"orster (solid), Leegwater (dotted), and quantum (symbols) average transfer times for the $N=3$ linear chain.
Here we fix $\Omega =1$. On the horizontal axis we normalize the trapping by
$\Gamma_{\text{trap}}^{\rm opt}$ as given by Eq.~(\ref{gopt}).
}
\label{fig:1}
\end{figure}

First, results for $N=3$ sites are reported in Fig.~\ref{fig:1}.   
We see that in the case $W \gg \gamma \gg \Omega$ ($W=100$, $\gamma=10$), 
the exact quantum results agree very well with the Leegwater approximation, 
and the maximal quantum enhancement occurs at $\Gamma_{\text{trap}} \approx \Gamma_{\text{trap}}^{\rm opt}$,
as predicted. For $W \sim \gamma$ (see $W=3, \gamma=1$), quantum transport enhancement becomes negligible, and the enhancement effect disappears entirely for $W \lesssim \gamma$ (see $W=0.3, \gamma=0.3$). Where a noticeable difference is observed between the Leegwater approximation and the exact quantum results, the exact quantum corrections favor somewhat greater coherent transport enhancement, i.e.,  the true enhancement is slightly stronger than that predicted by the Leegwater model (see for example $W=10, \gamma=1$ in Fig.~\ref{fig:1}). This correction is addressed at a quantitative level below.

\begin{figure}
\includegraphics[width=11cm]{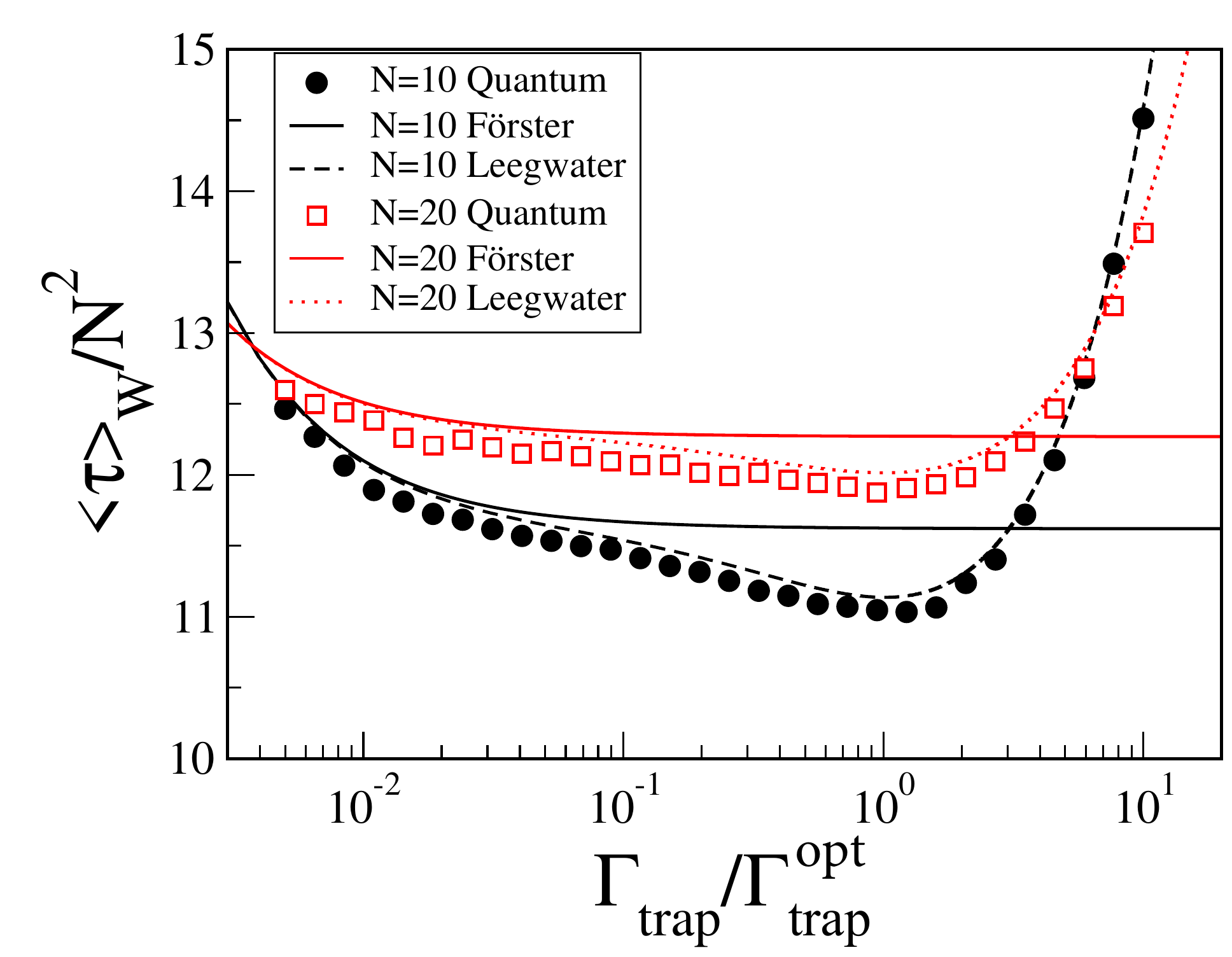}
\caption{ (Color online)
Comparison among F\"orster (solid cures), Leegwater (dotted and dashed curves),  and  quantum (symbols) 
average transfer times (rescaled by  a factor $N^2$) for a linear chain with $N=10$ and $N=20$ sites.
Here we fix $\Omega =1$, $\gamma=10$, and $W=50$.
 On the horizontal axis we renormalize the trapping by $\Gamma_{\text{trap}}^{\rm opt}$ (Eq.~(\ref{gopt})). 
}
\label{fig:2}
\end{figure}

Next, we confirm that these results continue to hold for long chains ($N \gg 1$). In Fig.~\ref{fig:2}. The results of the F\"orster
model (\ref{cl}) and of the Leegwater approximation (\ref{leeg}) are compared with the exact quantum calculation for $N=10$ and $20$ a fixed set of parameters such that $W \gg \gamma \gg \Omega  $.
It is clear from  Fig.~\ref{fig:2} that the above picture continues to hold at large $N$: The  difference between incoherent and quantum transfer times is still maximized for $\Gamma_{\text{trap}} \simeq\Gamma_{\text{trap}}^{\rm opt}$, the minimal quantum transfer time
also occurs near $\Gamma_{\text{trap}}^{\rm opt}$,
and the Leegwater approximation underpredicts the true quantum enhancement by a slight margin.

\begin{figure}
am	\includegraphics[width=11cm]{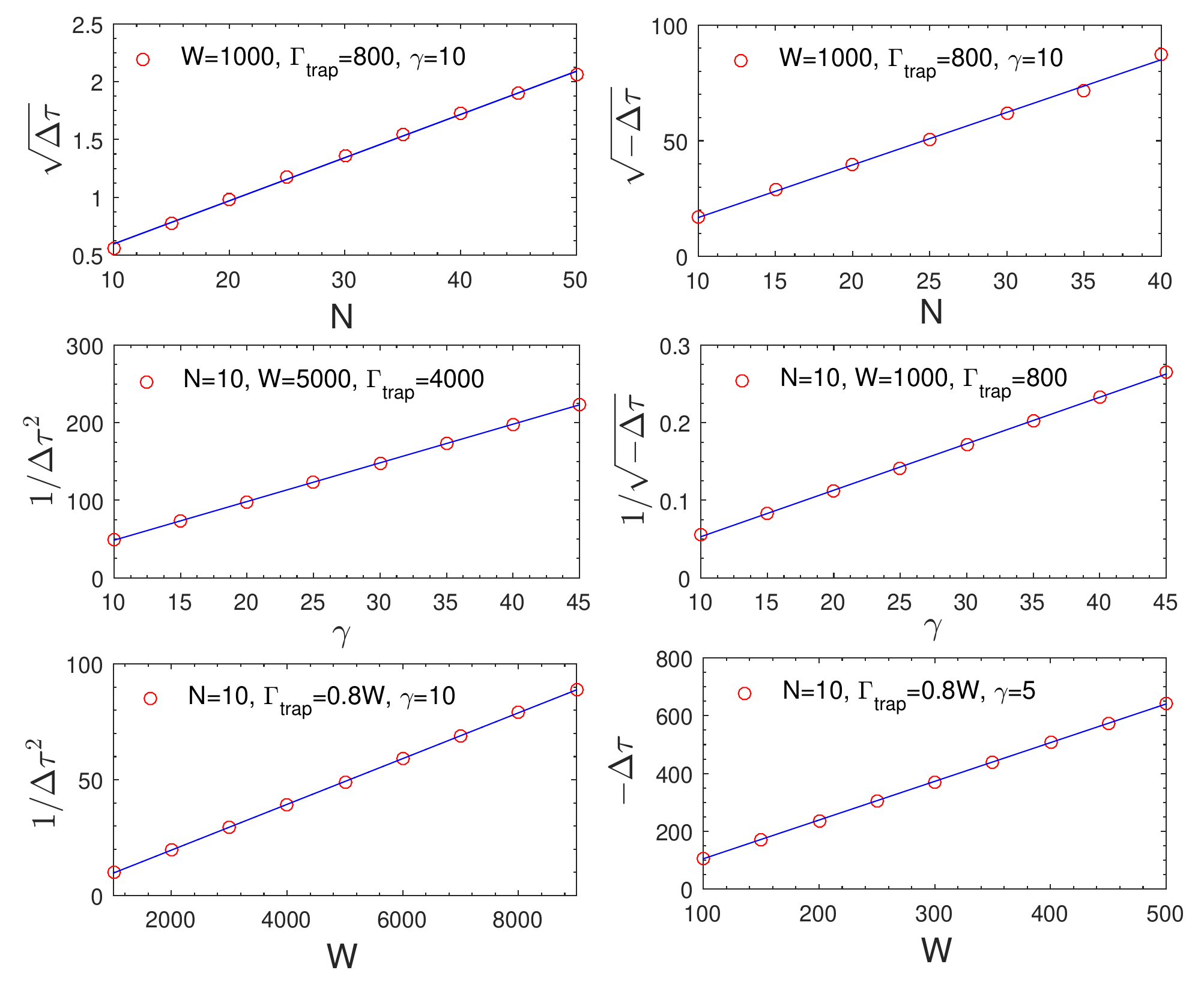}
	\caption{ (Color online)
		The error in the Leegwater approximation, $\Delta \tau=\left<\tau\right>-\left<\tau_{L}\right>$, is shown as a function of system size $N$, dephasing strength $\gamma$, and disorder strength $W$, in the parameter regime of interest. The panels on the left side are for the fully-connected network and those on the right side are for the linear chain. The solid blue lines in the two top panels are fits to $(a'+b'N)^2/\sqrt{\gamma W}$ for the fully-connected network (with $a'=2.33$, $b'=0.37$) and to $-(a+bN)^2 W/\gamma^2$ for the chain (with $a=1.82$, $b=0.72$), which converge to Eqs.~(\ref{fcerr}) and (\ref{linerr}), respectively in the large-$N$ limit. We fix $\Omega=1$, and the other parameters are shown in the legend of each panel. Specially we fix $\Gamma_{\rm trap}=0.8 W$, which is close to the optimal opening $\Gamma_{\rm trap}^{\rm opt}=\sqrt{6}W/3$ for both systems.  
	}
	\label{fig:diff}
\end{figure}

We now briefly return to the observation in Figs.~\ref{fig:1} and \ref{fig:2} of a slight discrepancy between the exact quantum calculation and the Leegwater expressions. Although no analytic expression is available for the quantum chain with $N>2$, a systematic numerical scaling analysis in
the regime of interest, $N \gg 1 $, $W \gg \gamma \gg \Omega$, shows that the leading quantum correction to the Leegwater formula scales as
\begin{equation} 
 \left<\tau\right>_W - \left<\tau_{L}\right>_W = - b N^2 W /\gamma^2 \,,
 \label{linerr}
\end{equation}
where $b \approx 0.72$ is a constant, see Fig.~\ref{fig:diff}. 
Comparing with Eq.~(\ref{leeg}), we find the relative error in the Leegwater approximation:
 \begin{equation} 
\left<\tau\right>_W  / \left<\tau_{\rm L}\right>_W =  1  - O ( \Omega^2/ W \gamma) \,,
 \end{equation}
 independent of chain length for $N \gg 1$. This agrees with our previous observations: the Leegwater approach provides an excellent approximation to the quantum transfer time in the regime where quantum enhancement is possible, and the leading correction favors even slightly faster transport than that predicted by the Leegwater formula.

\subsection{Heuristic derivation of transfer times for the linear chain}
Here we give an heuristic derivation of the
average transfer time obtained in the previous section. 
In particular we will
analyze  the parameter regime where quantum 
transport outperforms
 incoherent transport.

Consider a linear chain of $N$ sites. We start at one end
of the chain and evaluate  the probability to reach the other end where the
excitation can escape with a rate $\Gamma_{\rm trap}$.

Let us first compute the average transfer time in the F\"orster model.  
To go from site $1$ to site $2$ takes an average time $1/T_{\rm F}$. The total
time required to perform the random walk from site $1$ to site $N$ scales as $N^2$,
or more precisely, $N(N-1)/T_{\rm F}$. 
Moreover, if the probability to be at the $N$-th site is 
$1/N$ and the escape rate is  $\Gamma_{\rm trap}$ we can estimate the exit
time as $N/\Gamma_{\rm trap}$.  Adding up the diffusion time and the exit time we
have,
\begin{equation}
\tau_{\rm F}= \frac{N(N-1)}{2 T_{\rm F}} + \frac{N}{\Gamma_{\rm trap}}= \frac{N}{\Gamma_{\rm trap}} + N(N-1)
 \left(
  \frac{6\gamma^2 +W^2}{24
    \Omega^2 \gamma} \right) \,,
\label{H1}
\end{equation}
which is exactly the result found by direct calculation, see
Eq.~(\ref{cl}). 
On the other hand, in the presence of an opening, the Leegwater formulas are
modified by the substitution $\gamma \rightarrow \gamma+ \Gamma_{\rm trap}/2$ for
the transfer rate between the last two sites (since in a linear chain only site $N-1$
is connected to the $N$-th site where the excitation can escape).
Needless to say, while 
 for $\Gamma_{\rm trap} \ll \gamma$ the transfer rate
in presence of the opening  reduces to
the  incoherent one,   for $\Gamma_{\rm trap} \approx \gamma$ the
two rates are very different. In particular the rate is maximal for
$\gamma+\Gamma_{\rm trap}/2=W/\sqrt{6}$. 

Thus, the (coherent) effects induced by the opening can be included in an 
(incoherent) model of diffusion using the Leegwater expression. 
We can estimate the transfer time for the quantum case in a similar
way as was done above, namely:  
\begin{equation}
\tau_{\rm L}= \frac{N}{\Gamma_{\rm trap}} + \frac{(N-1)(N-2)}{2 T_{\rm F}} +
\frac{N-1}{T_{\rm L}} \,.
\label{H2}
\end{equation}
This expression, rearranged, is the same as Eq.~(\ref{leeg}).
From the above expression we get,
\begin{equation}
\tau_{\rm F}-\tau_{\rm L}= (N-1)\left(\frac{1}{T_{\rm F}}-\frac{1}{T_{\rm L}}\right) \,.
\label{H3}
\end{equation}
This last expression is simpler to analyze: quantum transport is better
than incoherent transport  when $T_{\rm L}>T_{\rm F}$,
from which we have:
\begin{equation}
\Gamma_{\rm trap} < \frac{W^2-6\gamma^2}{3\gamma}
\label{H4}
\end{equation}
which can be achieved only if $W>\sqrt{6}\gamma$. 
Moreover for $W \gg \Omega$ and $\Gamma_{\rm trap}>\gamma$, the optimal quantum transport is
obtained for $\Gamma_{\rm trap}=\Gamma_{\text{trap}}^{\rm opt}\equiv \sqrt{2/3}W-2\gamma$ (the same value that maximizes
 the rate $T_{\rm L}$). We can now compare the optimal quantum transport with the
optimal F\"orster transport obtained for  $\Gamma_{\rm trap}\to\infty$. 
So we have:
\begin{equation}
\tau_{\rm F}^{\rm opt} =  \frac{N(N-1)}{2T_{\rm F}}, 
\label{H5}
\end{equation}
\begin{equation}
\tau_{\rm L}^{\rm opt}= \frac{(N-1)(N-2)}{T_{\rm F}} + \frac{N-1}{T_{\rm L}^{\rm opt}}
+\frac{N}{\Gamma_{\text{trap}}^{\rm opt}}
\label{H6}
\end{equation}
and for $W \gg \gamma$, $N \gg 1$,
\begin{equation}
\tau_{\rm F}^{\rm opt}-\tau_{\rm L}^{\rm opt} \approx
 \frac{N W^2}{ 12\Omega^2 \gamma} \,.
\label{H7}
\end{equation}

Note that that quantum enhancement due to the opening is proportional
to the variance of the static disorder.

\section{Fully connected networks} \label{sec-fc}

In Sec.~\ref{sec-chain} we saw that coherent effects can aid transport through an open linear chain of arbitrary length, as long as the static disorder is sufficiently strong relative to the dephasing rate. To demonstrate the generality of this effect, we now consider a quantum network which, in its degree of connectivity, may be considered to be at the opposite extreme from a linear chain, namely a fully connected network with equal couplings between all pairs of sites, as illustrated in Fig.~\ref{network} (lower panel). Specifically, the Hamiltonian $\text{H}_{\text{lin}}$ (Eq.~(\ref{linham})) for the linear chain is replaced by 
\begin{equation}
\text{H}_{\text{fc}}=\sum_{i=1}^N  \omega_{i} \ket{i}\bra{i}+
\Omega \sum_{1 \le i <j \le N}  \left( \ket{i}\bra{j}+ \ket{j}\bra{i} \right)\,,
\end{equation}
and the site energies $\omega_i$ are again distributed uniformly in $[-W/2,W/2]$.
As before, site $N$ is coupled to the continuum with decay rate $\Gamma_{\rm trap}$ (Eq.~(\ref{amef})).
In the F\"orster model, then, every site is connected to every other with the incoherent rate $T_{\rm F}$ (Eq.~(\ref{Lcl}), where $\Delta$ is the difference between the site energies), whereas
in the Leegwater approximation the transfer rates $T_{j \to N}$ and $T_{N \to j}$ are given by the modified rate $T_{\rm L}$ (Eq.~(\ref{Lqu})), which includes the effect of the opening.
To simplify the analysis, we focus only on the regime where quantum transport enhancement is most pronounced: This occurs when the opening is comparable to the disorder strength, and the mean level spacing $W/N$ is large compared to the dephasing rate,
$\Gamma_{\rm trap}/N \sim W/N  \gg \gamma \gg \Omega$. In that case we have $T_{\rm F} \sim \Omega^2  \gamma/W^2$ and $T_{\rm L} \sim \Omega^2/W$, so $T_{\rm L} \gg T_{\rm F}$ and to leading order all sites are effectively coupled to site $N$ only. In this regime, the average time to reach the sink starting from site $1$ attains the $N$-independent value
\begin{equation}
\left<\tau_{\rm L} \right>_W = \frac{3\Gamma_{\rm trap}^2+2W^2}{12\Omega^2\Gamma_{\rm trap}} \,.
\label{fcleeg}
\end{equation}
We note that this expression agrees, as it must, with the linear chain result (\ref{leeg}) for the case $N=2$ in the limit $\Gamma_{\rm trap} \sim W  \gg \gamma \gg \Omega$. The
transfer time is minimized,
\begin{equation}
\left<\tau_{\rm L} \right>_W^{\rm min}=W/\sqrt{6}\Omega^2 \,,
\label{taumin}
\end{equation}
when the opening strength is set to the optimal value
\begin{equation}
\Gamma_{\text{trap}}^{\rm opt}=\sqrt{2/3} W \,.
\label{goptfc}
\end{equation}

The above discussion addresses the transfer time in the context of the Leegwater approximation. As in the case of the linear chain, an exact numerical evaluation of the average quantum transfer time confirms that the Leegwater approximation provides the leading contribution to the quantum transfer time in the regime of interest. The leading correction for the error takes the form 
\begin{equation}
\left<\tau\right>_W-\left<\tau_{\rm L} \right>_W  \approx (b'N)^2/\sqrt{\gamma W} \,,
\label{fcerr}
\end{equation} 
with $b' \approx 0.37$, as illustrated in Fig.~\ref{fig:diff},
so
  \begin{equation} 
  \left<\tau\right>_W  / \left<\tau_{\rm L}\right>_W =  1  + O ( N^2 \Omega^2/ W^{3/2} \gamma^{1/2}) \,.
  \end{equation}

\begin{figure}
	\includegraphics[width=10cm]{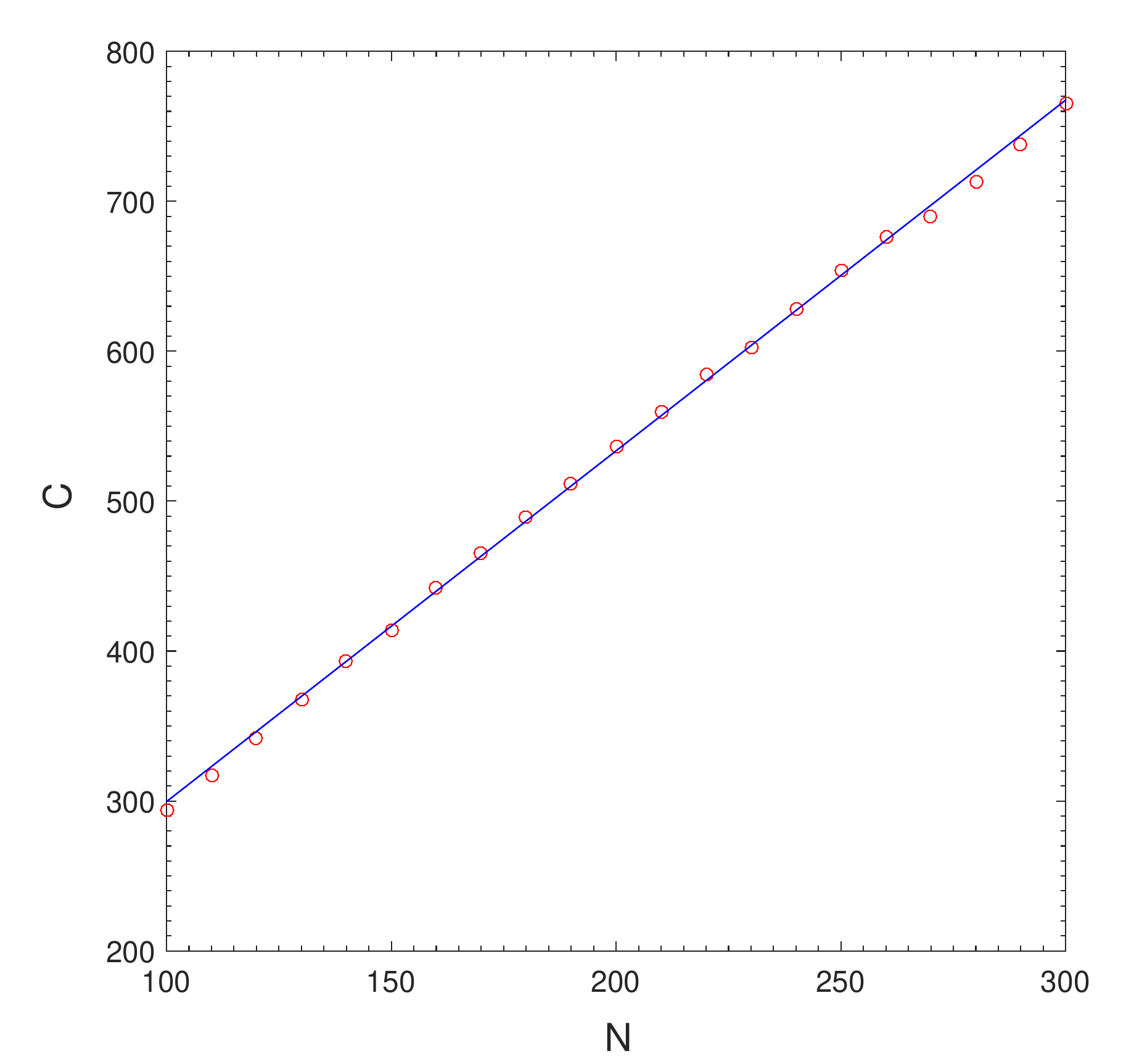}
	\caption{(Color online) The function $C(N)$ in Eq.~(\ref{fullclasnum}), which describes the $N$-dependence of the incoherent transfer time in the fully connected model, with a fit to $C(N)=2.34 N +64.55$. Here $W=5000$, $\Gamma_{\rm trap}=6000$, and $\gamma=\Omega=1$. 
	}
\end{figure}

To obtain the corresponding F\"orster behavior, it is convenient to work in the large-$N$ limit. The probability to jump from site $i$ to site $j$
is given by the F\"orster transition rate (\ref{Lcl}),
\begin{equation}
(T_{\rm F})_{i \to j}=\frac{2\Omega^2 \gamma}{\gamma^2+(\omega_i-\omega_j)^2}\,.
\end{equation}
Now if we label the sites in order of site energy, $\omega_1 < \omega_2 < \ldots  < \omega_N$, for large $N$ we have $(\omega_i-\omega_j)^2  \approx W^2(i-j)^2/N^2$.
Since we are working in the
regime of very strong disorder, $W \gg N \gamma$, the transition rates simplify to $(T_{\rm F})_{i \to j} \approx 2 N^2 \gamma \Omega^2/ W^2(i-j)^2$. This corresponds to an $\alpha=1$ L\'evy flight (or Cauchy flight) with typical time scale $\Delta t \sim W^2/N^2 \gamma \Omega^2$ for each jump; for an $\alpha=1$ L\'evy flight the average time to travel a distance $n$ scales as $n$, in contrast with the $n^2$ scaling of the travel time for ordinary diffusion~\cite{levy}. Although the initial site is not necessarily site 1 due to the site relabeling, and the site coupled to the sink is not necessarily site $N$, the initial and final sites are nevertheless separated by a distance of order $N$. Thus, total time required to travel through the system scales as $\tau \sim N \Delta t$, and we have
\begin{equation}
\left< \tau_{\rm F} \right>_W  \sim \frac{W^2}{N\gamma\Omega^2} \,. 
\label{fullclas}
\end{equation} 
The behavior given in Eq.~(\ref{fullclas}) is confirmed by exact numerical calculations. Numerically we obtain an excellent fit to
 \begin{equation}
 \left< \tau_{\rm F} \right>_W  \approx \frac{W^2}{C(N)\gamma\Omega^2} \,,
 \label{fullclasnum}
 \end{equation} 
where $C(N)\approx 2.34 N +64.55$. We note that from Eq.~(\ref{fullclas}) or Eq.~(\ref{fullclasnum}) the incoherent transfer time may appear to approach $0$ in the large-$N$ limit; however
one must keep in mind that the above discussion assumes $W/N \gg \gamma$. If we increase $N$ while holding all other system parameters fixed, we find instead that for $N > W/\gamma$, the F\"orster transfer time saturates at an $N$-independent value $\left< \tau_{\rm F} \right>_W \sim W/ \Omega^2$, comparable to the Leegwater prediction. In this limit there is no significant quantum
enhancement of transport.

\begin{figure}
	\includegraphics[width=17cm]{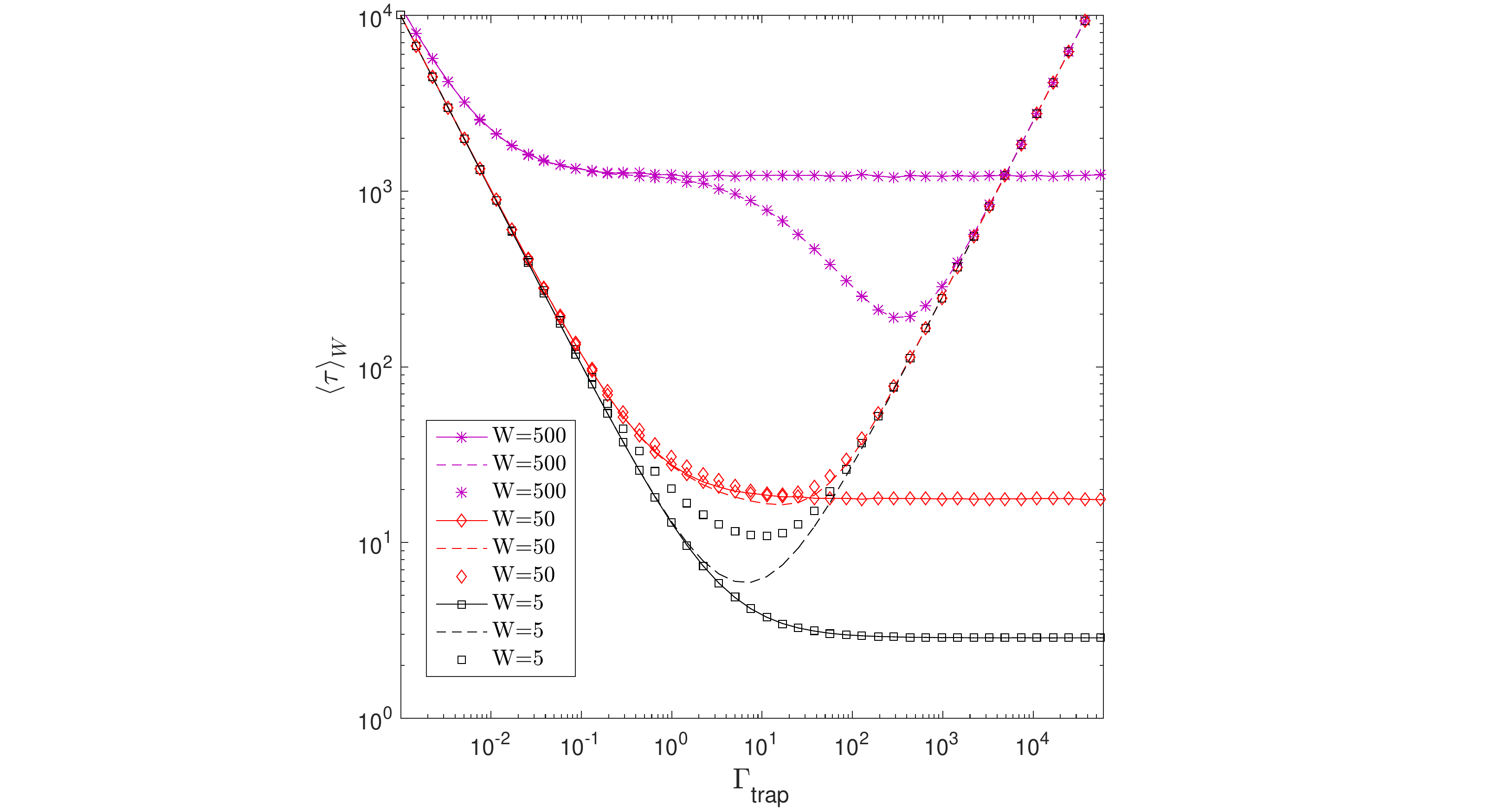}
	\caption{(Color online) Comparison among F\"orster (solid curves with symbols), Leegwater (dashed cruves),  and  quantum (symbols only) 
		average transfer times for a fully connected network with $N=10$  sites and several values of the disorder strength $W$.
		Here we fix $\Omega =1$ and $\gamma=5$. 
	}
	\label{fig:fcn10}
\end{figure}

Returning to the regime of primary interest, $W/N  \gg \gamma \gg \Omega$ and comparing Eqs.~(\ref{taumin}) and (\ref{fullclas}) we find a very strong coherent enhancement of transport in the fully connected network. Specifically, when the opening strength $\Gamma_{\rm trap}$ is of order $\Gamma_{\text{trap}}^{\rm opt}$, the ratio of the Leegwater (or, equivalently, quantum) transfer time to the incoherent time scales as
 \begin{equation}
 \frac{\left< \tau_{\rm L} \right>_W }{\left< \tau_{\rm F} \right>_W} \sim  \frac{\gamma}{W/N} \ll 1 \,.
 \end{equation}
We note that the condition $W/N \gg \gamma$ which allows quantum mechanics to significantly aid transport in the fully connected network corresponds precisely to the starting assumption 
underlying the calculations in this Section.

\begin{figure}
	\includegraphics[width=13cm]{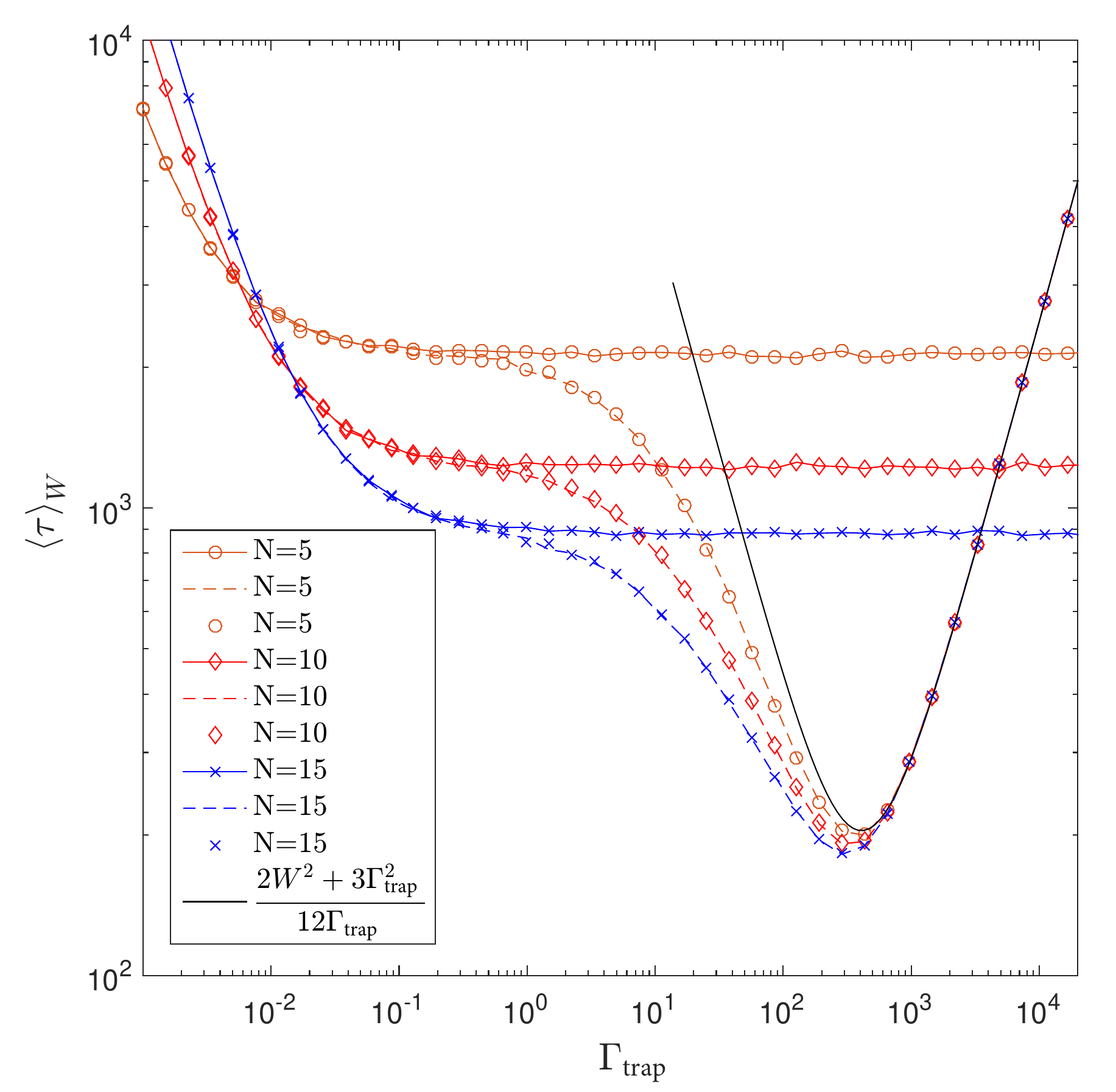}
	\caption{(Color online) Comparison among F\"orster (solid curves with symbols), Leegwater (dashed curves),  and  quantum (symbols only)
		average transfer times for fully connected networks of several sizes $N$.
		Here we fix $W=500$, $\gamma=5$, and $\Omega =1$. The analytic result (\ref{fcleeg}) describes the behavior of the Leegwater and quantum ensemble-averaged transfer time in the region
		of strongest quantum transport enhancement. 
	}
	\label{fig:fc2}
\end{figure}

The results in Fig.~\ref{fig:fcn10} confirm that a very strong quantum enhancement of transport occurs in a fully connected network of $N=10$ sites for $W \gg N \gamma$ (see the data for $W=500$ in the figure). Optimal quantum transport appears at the value of the opening given by $\Gamma_{\text{trap}}^{\rm opt}$. We also observe excellent agreement between the exact quantum calculation and the Leegwater approximation in this regime. The Leegwater approximation breaks down for larger relative values of the dephasing rate $\gamma$ (e.g. $W=\gamma=5$), but in this range of parameters quantum effects hinder rather than aid transport. Fig.~\ref{fig:fc2} illustrates that in the region
of strongest quantum transport enhancement ($\Gamma_{\rm trap} \sim \Gamma_{\text{trap}}^{\rm opt}$), the quantum behavior is indeed approximately $N$-independent and is well described by the analytic expression given in Eq.~(\ref{fcleeg}).

\section{The FMO complex} \label{sec-fmo}

The FMO photosynthetic complex has received a lot of attention in recent years as an example of a biological system that exhibits
quantum coherence effects even at room temperature~\cite{photo,photoT,photo2}.
In particular, the interplay of opening and noise in the FMO complex has been already analyzed in Ref.~\cite{srfmo}, where it was shown that even at room temperature, the superradiance transition is able to enhance transport.
Here we examine opening-assisted quantum transport enhancement in the FMO complex and observe that the same behavior obtains here as in the linear chain and fully connected model systems considered in the previous sections.
\begin{figure}
	\includegraphics[width=9cm]{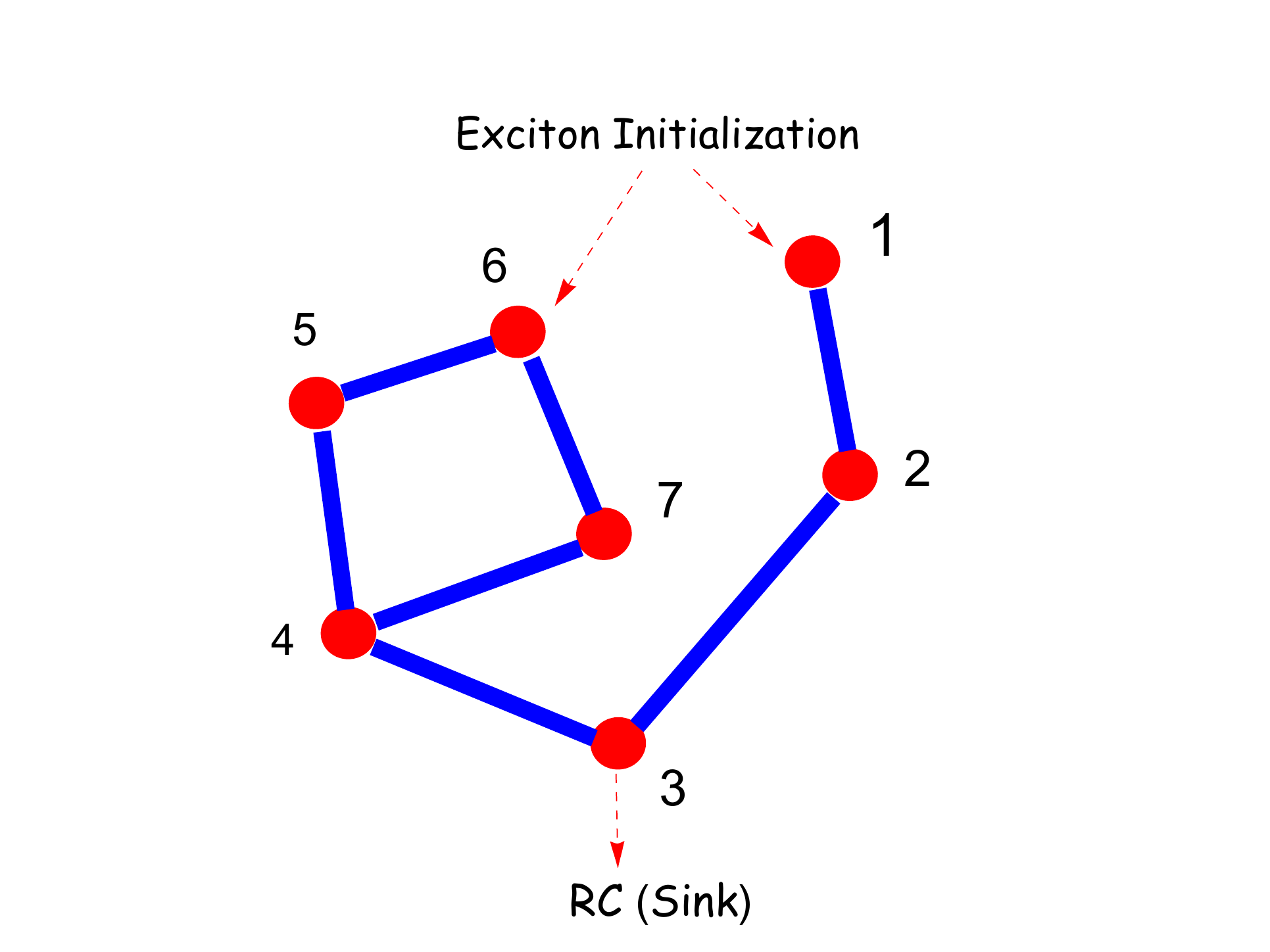}
	\caption{(Color online) A schematic illustration of the FMO Hamiltonian (\ref{hfmo}), with each bond indicating a coupling matrix element of magnitude at least $20~{\rm cm}^{-1}$.
	}
	\label{fig:fmo-schem}
\end{figure}

Each subunit of the FMO complex contains seven chromophores, and may be modeled by the tight-binding Hamiltonian
 \begin{equation}
\text{H}_{\rm FMO}= \left(\begin{matrix} 
200  & -87.7 & 5.5  & -5.9 & 6.7 & -13.7 & -9.9  \\
-87.7  & 320 & 30.8  & 8.2 & 0.7 & 11.8 & 4.3  \\
5.5  & 30.8 & 0  & -53.5 & -2.2 & -9.6 & 6 \\
-5.9  & 8.2 & -53.5  & 110 & -70.7 & -17 & -63.3 \\
6.7  & 0.7 & -2.2  & -70.7 & 270 & 81.1 & -1.3  \\
-13.7  & 11.8 & -9.6  & -17 & 81.1 & 420 & 39.7 \\
-9.9  & 4.3 & 6  & -63.3 & -1.3 & 39.7 & 230  
\end{matrix}  \right) {\rm cm}^{-1} \,,
\label{hfmo}
 \end{equation}
in units where $hc=1$.  We notice that the connectivity between the sites is greater than that in a linear chain, but the
 inter-site couplings are very non-uniform. Thus, this realistic system may be considered to be intermediate between a chain and a fully connected network. A schematic illustration of the Hamiltonian $\text{H}_{\rm FMO}$, where only off-diagonal elements of magnitude greater than $20~{\rm cm}^{-1}$ are indicated by bonds, appears in Fig.~\ref{fig:fmo-schem} (but in all calculations below we employ the full Hamiltonian given in Eq.~(\ref{hfmo})).
 
 Since incident photons are believed to create excitations on sites 1 and 6 of the FMO complex~\cite{lloyd}, we take the initial state of the system to be
 \begin{equation}
 \rho(0)=\frac{1}{2}\left(|1\rangle\langle 1| +|6\rangle\langle 6|\right) \,.
 \end{equation}
 
 Site $3$ is coupled to the reaction center, which serves as the sink for the FMO complex, with decay rate $\Gamma_{\rm trap}$. Additionally, an excitation on any site may decay through exciton recombination with rate 
 $\Gamma_{\rm fl}=(1~{\rm ns})^{-1}=0.033~{\rm cm}^{-1}$~\cite{lloyd,deph}, but this slow decay has a negligible effect on the transfer time $\tau$, as discussed in Sec.~\ref{sec:2}.
 
 \begin{figure}
 	\includegraphics[width=13cm]{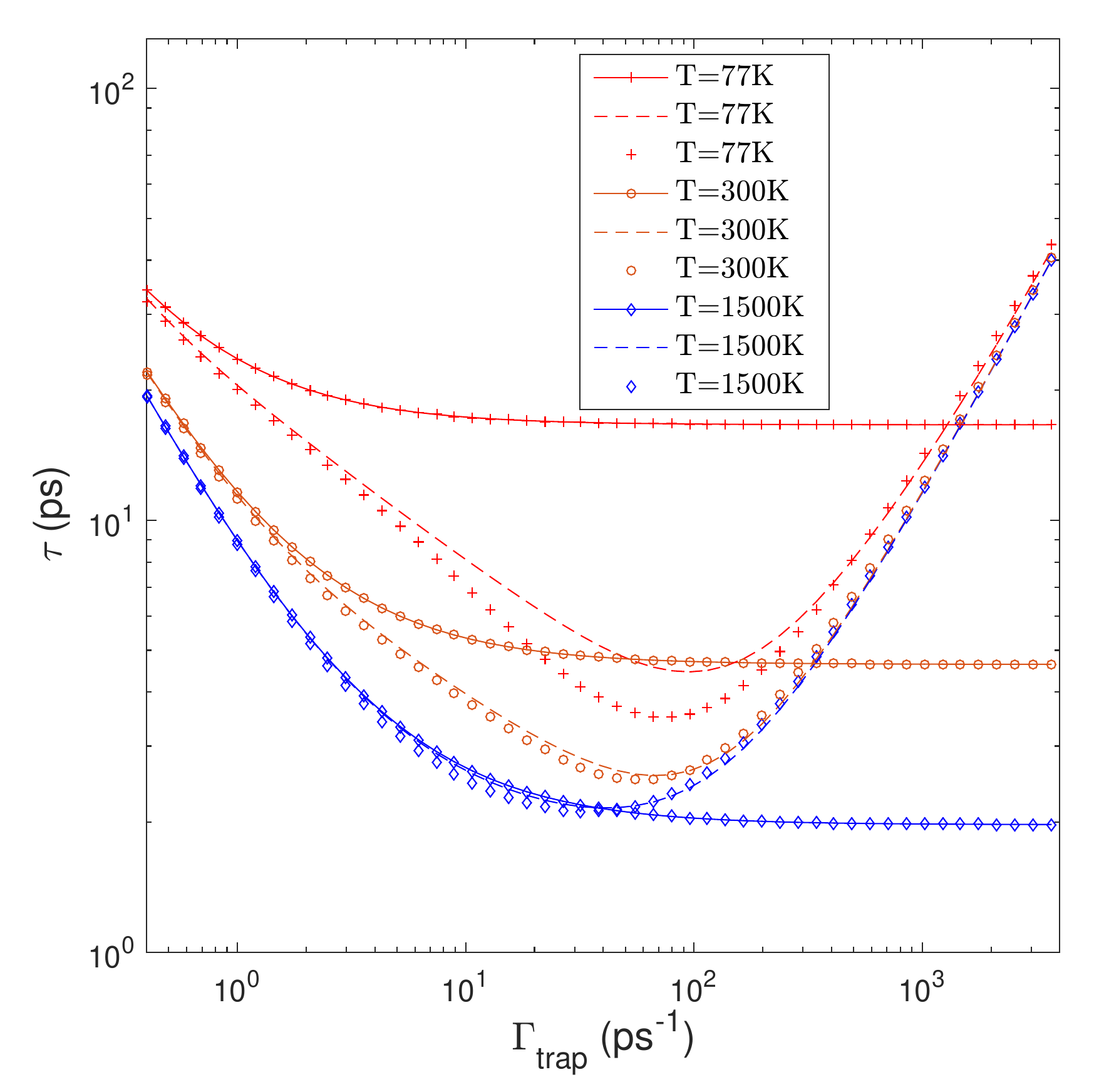}
 	\caption{(Color online) The average transfer time for the FMO system is shown as a function of coupling $\Gamma_{\rm trap}$ between site 3 and the reaction center, at three different temperatures. The  F\"orster, Leegwater,  and  quantum transfer times are represented by solid curves with symbols, dashed curves, and symbols, respectively.
 	}
 	\label{fig:fmo-1}
 \end{figure}
 
 \begin{figure}
 	\includegraphics[width=13cm]{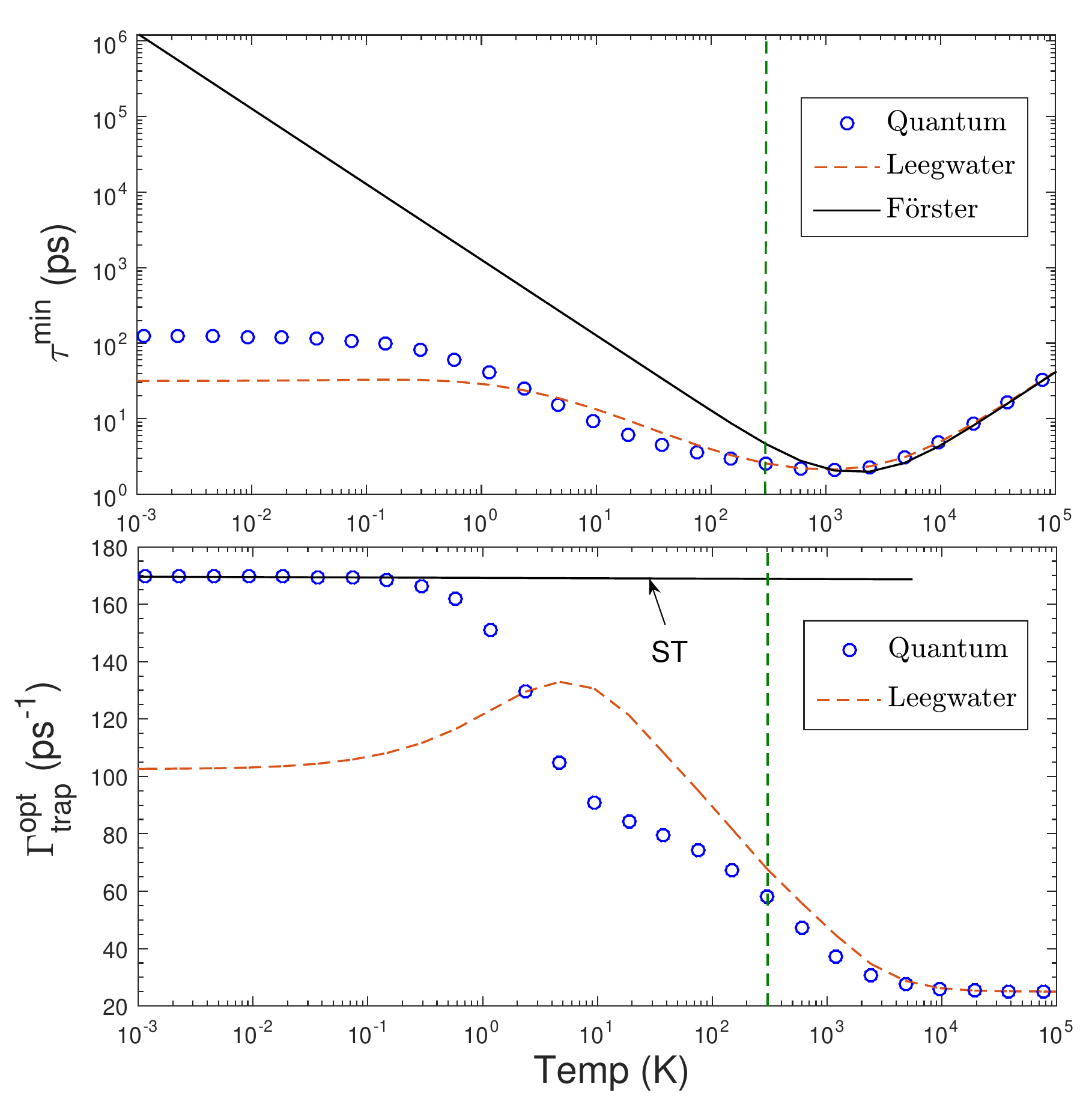}
 	\caption{(Color online) Upper panel: The minimal transfer time through the FMO complex (optimizing over the coupling $\Gamma_{\rm trap}$ to the reaction center) is shown as a function of temperature, for the quantum, Leegwater, and incoherent (F\"orster) calculations. Lower panel: The optimal coupling $\Gamma_{\rm trap}^{\rm opt}$ is shown as a function of temperature, in the full quantum calculation and in the Leegwater approximation. In the incoherent model, the optimal coupling is always $\Gamma_{\rm trap}^{\rm opt}=\infty$. The horizontal solid line indicates the location of the superradiance transition at zero temperature. In both panels, the dashed vertical line indicates room temperature, $T=300$~K. 
 	}
 	\label{fig:fmo-2}
 \end{figure}

 The transfer time calculation as a function of reaction center coupling $\Gamma_{\rm trap}$ is shown in Fig.~\ref{fig:fmo-1} (see also Ref.~\cite{srfmo}). For the FMO system, the dephasing rate $\gamma$ is related to the temperature by the relation $\gamma = 0.52c(T/{\rm K})~{\rm cm}^{-1}$, where $T/{\rm K}$ is the temperature in Kelvin units~\cite{photoT}, and results for three values of the temperature (or equivalently, dephasing rate) are shown in the figure. Notably, strong opening-assisted quantum enhancement of transport is seen not only at liquid nitrogen temperature (77 K) but also at room temperature (300 K) where the quantum transfer is up to a factor of 2 faster than that obtained by an incoherent calculation. We also see good agreement between the exact quantum calculation and the Leegwater approximation at room temperature. For comparison, we show an example at very high temperature (1500 K), where the quantum transport enhancement is almost absent.
 
 Although the ``disorder'' in the FMO Hamiltonian is fixed, for the purpose of estimating the relevant energy, time, and temperature scales we may analogize this Hamiltonian to one drawn from a disordered ensemble. The variance of the site energies $(H_{\rm FMO})_{ii}$ is $\sigma^2=(128~{\rm cm}^{-1})^2$,  which corresponds to $W=443~{\rm cm}^{-1}$. Then we see that room temperature, $T=300~$K or $\gamma=156~{\rm cm}^{-1}$, actually corresponds to a marginal case where ``static disorder'' $W$ and dephasing rate $\gamma$ are comparable. At even higher (biologically unrealistic) temperatures, e.g. $T=1500$~K, we have $\gamma=780~{\rm cm}^{-1} \gg W$, and quantum enhancement of transport is absent, as expected. At lower (also unrealistic) temperatures, e.g. $T=77$~K, we have $\gamma=40~{\rm cm}^{-1} \ll W$, corresponding to a regime where opening-assisted quantum transport enhancement is most pronounced. The crossover between the low-temperature regime where coherent effects strongly aid transport and the high-temperature regime where coherent effects provide no advantage is studied quantitatively in Fig.~\ref{fig:fmo-2} (upper panel), where the minimal quantum, Leegwater, and F\"orster transfer times are shown at each temperature (optimizing in each case over the opening strength $\Gamma_{\rm trap}$).
 
 Similarly we may estimate the optimal strength of the opening at low temperature using the formula $\Gamma_{\text{trap}}^{\rm opt} =\sqrt{2/3}W$ obtained for the linear chain and fully connected network at small dephasing and $\Omega \to 0$ (see Eqs.~(\ref{gopt}) and (\ref{goptfc})). This gives $\Gamma_{\text{trap}}^{\rm opt}=\sqrt{2/3}(2\pi c)(443~{\rm cm}^{-1})=68~{\rm ps}^{-1}$, which is in reasonable qualitative agreement with the location of the Leegwater and quantum minima at liquid nitrogen temperature in Fig.~\ref{fig:fmo-1}. (The above formula is valid for inter-site couping $\Omega \to 0$, and therefore is expected to underestimate the true value of $\Gamma_{\text{trap}}^{\rm opt}$). As expected from our study of the two-site model and linear chain (see Eqs.~(\ref{Goptsmg}) and (\ref{gopt})), the location of the minimum shifts to smaller coupling $\Gamma_{\rm trap}$ as the temperature (dephasing) increases. The full dependence of the optimal opening strength on temperature in the exact quantum calculation as well as in the Leegwater approximation are shown in detail in Fig.~\ref{fig:fmo-2} (lower panel).

\section{Conclusions}\label{sec:6}

We have analyzed the role of the opening in enhancing coherent transport in the presence of both disorder and dephasing. The effect is investigated in several
paradigmatic models, including a two-site system, a linear chain of arbitrary length, and a fully connected network of arbitrary size. For the two-site model, fully analytical expressions exist for both the incoherent and quantum average transfer times, and therefore the regime in which coherent effects aid transport as well as the optimal opening strength at which the effect is maximized may also be obtained analytically. For the linear chain and fully connected network, we are able to find analytical expressions
in the deep classical regime, where dephasing is much stronger than
the hopping coupling between the sites. In this case quantum transport can be
described with an incoherent master equation where the rates incorporate the effect of the opening, as suggested by Leegwater. Again, the different efficiencies of quantum and incoherent transport can be compared to identify the regime in which coherent effects aid transport. In this regime we find the
optimal opening able to maximize transport efficiency. We see very generally that quantum transport can outperform incoherent transport even at high rates of dephasing (or dynamic disorder), as long as the static disorder strength is sufficiently large. The optimal strength of the opening grows linearly with the disorder strength. 
An analysis of the FMO
natural photosynthetic complex 
confirms the role of the opening in enhancing coherent transport in realistic models, even at room temperature.

\acknowledgments

This work was supported
in part by the NSF under Grant No. PHY-1205788 and by the Louisiana Board of Regents under contract LEQSF-EPS(2014)-PFUND-376.

\end{document}